\documentclass{emulateapj}
\usepackage{graphicx}
\usepackage{multirow,natbib}
\usepackage{textcomp}
\usepackage{epsfig}
\usepackage{amsmath}
\usepackage{amssymb}
\usepackage{amsthm}
\usepackage{hhline}
\usepackage{xy}
\usepackage{pbox}

\begin{document}

\title{Time and energy dependent characteristics of thermonuclear burst oscillations}
\author{Manoneeta Chakraborty$^{1,2}$ $^*$ ,  Yunus Emre Bahar$^{1}$, Ersin  G\"o\u{g}\"u\c{s}$^{1}$}

\affil{$^{1}$ Sabanc\i~ University, Faculty of Engineering and Natural 
  Sciences, Orhanl\i ~Tuzla 34956 Istanbul Turkey \\
  $^{2}$ Center of Astronomy, Indian Institute of Technology Indore, Khandwa Road, Simrol
Indore 453552, India}
  
\email{$^*$ manoneeta@sabanciuniv.edu}

\begin{abstract}\label{Abstract}
We have investigated temporal and spectral properties of a large sample of thermonuclear bursts with oscillations from eight different sources with spin frequencies varying from 270 to 620 Hz. For our sample we chose those bursts, for which the oscillation is sufficiently strong and of relatively long duration. The emission from the hot-spot that is formed during a thermonuclear burst is modulated by several physical processes and the burst oscillation profiles unavoidably carry signatures of these. In order to probe these mechanisms, we examined the amplitude and phase lags of the burst oscillations with energy. We also studied the frequency variation of oscillations during these thermonuclear bursts. We observed that the frequency drifts are more frequent in the cases where the spin frequency is lower. We found that the phase lag of the burst oscillations shows no systematic evolution with energy between the bursts, and also in between different sources. In 7 cases, we do indeed observe lag of soft energy photons, while there are a significant number of cases for which hard lag or no lag is observed. 
\end{abstract}


\section{Introduction}\label{Introduction}

Thermonuclear bursts are observed from accreting neutron stars in low mass X-ray binaries (LMXBs) when accreted matter gets ignited on the stellar surface triggering an unstable thermonuclear reaction. Such bursts originating from the neutron star surface are the most promising candidates to probe the degenerate supra-nuclear matter at the stellar core. The properties of thermonuclear bursts depend on the underlying physical conditions present during the runaway thermonuclear processes on the stellar surface. Various factors, such as, the accretion rate, the accreted fuel, magnetic field, stellar spin affect the burst characteristics. During thermonuclear bursts, the observed highly coherent periodic intensity variations are attributed to asymmetries developed on the surface of the rapidly rotating neutron star. Such burst oscillations were first reported by \citet{Strohmayer1996} using observations with Rossi X-ray Timing Explorer (RXTE). The detection of burst oscillation at 401 Hz from the millisecond pulsar SAX J1808.4--3658 \citep{Chakrabarty2003} proved the burst oscillation frequency to be synonymous with the spin frequency of the neutron star. The detected frequencies of burst oscillations range from 11 to 620 Hz. Burst oscillations are detected in about 10\% of thermonuclear bursts and are observed during all three phases of the burst, namely the rise, peak and tail. The oscillation frequency occasionally drifts by 1-2 Hz and and asymptotically approaches the spin frequency \citep{Galloway2008, Watts2012}. 

The burst oscillations during the rising phase is expected to originate due to an expanding hot-spot on the stellar surface as the thermonuclear flame propagates from the point of ignition. However, the origin of burst oscillations during the tail is still highly debated, and various models have been proposed to explain them. The two most promising models for burst decay oscillations are the propagating surface wave modes in the neutron star ocean (see \citet{Heyl2004}, \citet{Watts2012} and references therein) and the asymmetric cooling wake model \citep{Mahmoodifar2016}.  

Burst oscillation also offers a promising tool to probe various questions about the emission mechanism, such as, the physics of the thermonuclear burning front propagation. The emission from the hot-spot giving rise to the asymmetry gets modulated by relativistic effects, in particular by Doppler shift and light bending. The emission profile of signal from the hot-spot gets distorted due to these effects and hence an accurate modeling of the oscillation waveform can put new constraints on the mass and radius of the neutron star \citep{Lo2013, Cadeau2005}. Doppler effect and the light bending are dependent on the mass and radius of the neutron star. A strong signature of the special relativistic Doppler effect is proposed to be the presence of lags in the oscillation profile at the softer energy bands \citep{Strohmayer1999, Ford1999, Strohmayer2000}. For extremely rapidly spinning LMXBs this effect might become very significant. Ford (1999) reported a soft lag in the burst oscillation detected from a thermonuclear burst from the source Aquila X-1. \citet{Muno2003} investigated the energy dependence of the amplitude and phase variation burst oscillation for eight sources. For non-pulsating weakly magnetized LMXBs, the oscillation exhibits negligible fraction of harmonic content and the profiles are nearly sinusoidal \citep{Muno2003, Bhattacharyya2005, Watts2012}. The relatively low amplitudes of these oscillations indicate ignition near the polar latitudes. By examining the amplitude and phase evolution of burst oscillation profiles, it was found that the averaged profiles exhibit phase variations inconsistent with the theoretical expectations \citep{Muno2003}. In certain cases, a monotonically increasing hard lag contrary to a soft one was reported (Figure 2 of \citet{Muno2003}). This fact lends support to the suggestion that in such cases the Doppler effect is less significant compared to other processes, such as, Comptonization which might give rise to the hard lags \citep{Ford1999}. If such a scenario is indeed the case, the method of constraining the neutron star equation of state through burst oscillation profile modeling becomes ineffective as other significant systematic effects would then come into play. \citet{Artigue2013} examined four of these bursts and a super-burst from 4U 1636--536 individually and inferred that the phase lag variation is fully consistent with a rotating hot-spot model: The phase of the burst oscillation either showed soft lag or even when it showed random variation it was consistent within statistical errors with their theoretically predicted behavior assuming a rotating hot-spot model. The amplitude of the oscillation was observed to display a monotonically increasing behavior with energy which was also consistent with expectations. The lack of the harmonic content indicated a relatively simpler sinusoidal emission mechanism possibly closer to the rotational poles \citep{Muno2002b}. 

\citet{Maurer2008} simulated burst temporal profiles as a function of the various ignition conditions. They inferred that depending on latitude where the ignition is triggered, the burst rise profile shows either a convex or a concave shape. They parametrized the burst rise profile by a figure-of-merit quantity termed convexity which is always positive for bursts ignited near the equator and can be both positive and negative for bursts ignited near the poles. The ignition latitude on the other hand is dependent on the accretion rate and on the consequent burning regimes that set in with the varying accretion rate \citep{Cooper2007}. The presence and characteristics of burst oscillations are dependent on the ignition latitude as various factors, such as, flame spreading, rotational modulation are responsible for preserving the brightness asymmetry. Therefore, it can be observed that the burst oscillation properties are correlated with the convexity parameter that quantifies the burst profile shape. \citet{Maurer2008} and \citet{Zhang2016} showed that the distribution of convexity values is different for bursts that exhibit oscillation as compared to those without oscillation.  This suggests that bursts ignited at certain ignition latitudes are more likely to show burst oscillations. 

Burst oscillations exhibit relatively simpler waveforms compared to the complex structured rotationally powered or magnetically powered pulse profiles. Although their frequency drift and their origin especially during the decay introduces certain uncertainty, they still have advantages over rotational powered pulsation as they are not modulated by the intricate and ambiguous emission processes near the magnetic pole. Thus burst oscillations are still ideal tools for constraining neutron star compactness through pulse profile modeling. Here, we investigated 50 thermonuclear X-ray bursts from eight neutron star LMXBs in detail, and obtained their time and energy dependent temporal and spectral characteristics. In particular, for each burst, we explored the evolution of burst oscillation frequency, oscillation amplitude, and phase of oscillations with energy, as well as temporal evolution of the spectral parameters and oscillation amplitudes in order to better understand their origin, duration and intrinsic factors governing them. We describe our analysis and present our results in \S~\ref{Observation}, and discuss the implications of these results in the final section.  

\section{Observation, data analysis and results}\label{Observation}

Our sample of bursts consists of 50 individual events from 4U 1636--536, 4U 1702--42, 4U 1728--34, KS 1731--260, MXB 1658--298, 4U 1608--52, SAX J1750.8--2900, 4U 1915--053 detected with RXTE between 1996 July 14 and 2005 September 17 \citet{Galloway2008}. We selected these LMXB sources having a range of spin frequencies to examine the effects of the rotation frequency on the properties of burst oscillation. For this investigation, we have made use of the data collected with the Proportional Counter Array (PCA) on-board RXTE. The PCA consisted of five proportional counter units (PCU), each of which comprises of one propane veto layer, three Xenon layers and one Xenon veto layer \citep{Jahoda2006}. The PCA offered good sensitivity in the range 2-60 keV and corresponded to an effective area of 6500 cm$^2$ (when all five PCUs were operational). We considered Event mode data (or GoodXenon mode data where present) as they offered time resolution down to $\mu$s scale, and sufficiently good energy resolution which are essential for our timing study of burst oscillations. We used HEASOFT v6.19 tools and CALDB version 20120110 to analyze the data.
 
We first searched for burst oscillations in these 50 thermonuclear X-ray bursts. We focused on the oscillations detected in the tail of thermonuclear bursts though for comparison we have included about 10 cases where oscillation have been detected during the burst rise. We present our sample of bursts in Table~\ref{bpartiming}, which lists the source of the burst, spin frequency of the underlying neutron star, time and identification number of RXTE observation. This table also provides the burst rise time and the duration along with the oscillation parameters. The burst duration is defined as the time interval between the time when the intensity  increases sharply during the rise and the time at which the burst intensity falls down to 10\% of the burst peak intensity. For each burst, we identified the peak time of the burst and searched for oscillation in 2 s segments following the peak, shifting the each segment by a sliding time window of 0.25 s. In each time segment we calculated the power as a function of frequency taking a trial frequency range $\pm 2$ Hz within the reported spin frequency of the star with frequency steps of 0.001 Hz. The power was calculated using the Z$^2$ distribution which is effectively same as the fast-Fourier transform but can be applied in the case of continuous unbinned data. The Z$^2$ is defined as 
\begin{equation}
Z^2 = \frac{2}{N} \sum_{m=1}^{n} [\{\sum_{j=1}^{N} cos(m\phi_j)\}^2 +\{\sum_{j=1}^{N} sin(m\phi_j)\}^2 ]
\end{equation} 
where N is the total number of photons, n is the total number of harmonics and $\phi_j$ is the phase of the j$^{\rm th}$ photon. We have taken the number of harmonics n to be 1, as generally significant harmonics are not detected for the case of burst oscillations in non-pulsating sources. The Z$^2$ vs. frequency distribution thus obtained was fitted with a Gaussian function whose centroid frequency corresponded to the local oscillation frequency of that particular time segment. A threshold criterion was used on the Gaussian amplitude to mark a detection which corresponded to the 3 $\sigma$ power. For the segments where oscillation was significantly detected, the frequency and its error were evaluated. This procedure was repeated for all time segments during the burst and the time evolution of the frequency of significant burst oscillation was computed. We then modeled the time evolution of the frequency of burst oscillation using a constant and a quadratic (an approximation of the chirp) function. 
The parameters of the constant and quadratic models, and the corresponding fit statistics were computed. For the cases where the reduced chi-square of the constant model was greater than 1.3 and that of the quadratic model was lower, we used the latter model to describe the frequency evolution. Otherwise a constant model was employed. Additionally, we recorded the time intervals within which the oscillation was found to be significant. In Table.~\ref{bpartiming}, we also present the burst peak intensity, the average intensity per PCU during time when oscillation was significant and the time duration of the significant oscillation. The last 10 data points in Table.~\ref{bpartiming} following the horizontal line represents the cases where burst oscillations have been detected in only the rising phase of the bursts. For the oscillations during rise, it can be observed from Table.~\ref{bpartiming} $T_{osc}$ is sometimes greater than burst rise time which may seem counter-intuitive but this is because $T_{osc}$ is searched in overlapping 2 s coarse timebins whereas the burst rise time which was calculated at a much finer time resolution of 0.125 s. This different resolution for computing the two quantities may result in above mismatch if the oscillation lasts till the very end of the burst rise.

Next, we proceeded to uncover the energy dependence of the properties of burst oscillations. For this purpose, we first evaluated the phase variations of burst oscillations with energy. We calculated the phase evolution in the 5 energy bands corresponding to the absolute PCU channel ranges of 5-12, 13-21, 22-33, 34-43 and 44-71. In each channel (i.e., energy) interval, the time stamps were extracted from event file and the corresponding phases were obtained using the frequency evolution model which was already determined. By coherently folding the data, the phase folded light curves or the burst oscillation profiles with 16 phase bins were obtained. The contribution of the persistent emission was subtracted from the phase folded profile. The profiles were then fitted with a sinusoidal model, namely $A+B\sin(\phi+C)$, and computed the parameters of the sinusoidal model over all five energy ranges. The phase of sinusoidal function over the entire energy range was taken as the reference phase. The difference of the phase of each sine function in each energy band from that in the reference band gives the phase lag in that particular energy band. The $1\sigma$ errors on the phases in each energy band were propagated to obtain the $1\sigma$ error on the phase difference in each energy band. The fractional amplitude of the oscillation in each energy band was calculated by computing the ratio of parameters B to A for each band.

We find that the phase lag behavior in energy is either constant or varies significantly (see Fig.~\ref{fig01}). In this case we have focused on each individual burst oscillation event as compared to averaged profile for a source as presented by \citet{Muno2003}. Even though we do indeed see in a few cases a soft lag (Fig.~\ref{fig01}) as predicted to be arising from Doppler effects by theoretical models, we observe in most cases hard lags or no lags. 
Taking the entire energy band as the reference phase band may introduce some bias since the errors are not entirely independent in this case. To cross-check our results, instead of taking the entire energy range; we took the first band as the reference band and recomputed the phase lag variations. In this case, the errors on the phase lag values get comparatively larger given less number of counts was accumulated. We then compared the phase lag results obtained from these two methods and found that they completely agree with each other within error bars and the phase lag trends are the same in both cases (Fig.~\ref{fig1a}). 
Therefore the choice of the entire energy band as the reference phase band is justified considering statistical errors and we continue with this choice in this analysis.  
To quantify the phase lag evolution, we modeled the variation using a constant and a linear model. The parameters of the fits for our sample are presented in Table.~\ref{bparslope}. In  ~50\% cases, the linear model showed significant improvement (at least 50\%) of the reduced chi-square statistic over the constant model implying a lag between the energy bands. A negative slope corresponded to a soft lag whereas a positive slope implied a hard lag for the cases where the linear model was a better fit. Due to the small number of data points we do not quote any comparison test statistics between two models. There are some cases where neither of the models produce a good fit, which implies that in those cases the phase lag shows neither a constant nor a monotonic behavior. The fractional amplitude variation of the oscillation with energy showed usually a monotonic increasing trend (see Fig.~\ref{fig02}). There are a few cases where the fractional amplitude variation shows a fluctuating behavior but this can be mostly attributed to relatively weak burst oscillation signals.

We followed the same procedure as described above to compute the total fractional amplitude and considered the entire energy range in this case. The extracted time stamps within this range is folded according to the frequency model computed. The folded light curve is then again modeled using a sine function as before, and the corresponding fractional amplitude was calculated. We quote the total fractional amplitude for each burst showing oscillation in Table~\ref{bpartiming}. \citet{Mahmoodifar2016} put forward cooling wake models for explaining the origin of oscillations during the tail of the bursts. There are two categories of cooling wake models; the canonical cooling and the asymmetric cooling models. The canonical model predicts a correlation between the burst rise duration and the total burst tail oscillation amplitude. Our results show there is no such correlation present as shown by Fig.~\ref{fig3}, arguing against the canonical wake models and favoring the asymmetric cooling scenario. The correlation coefficient obtained between the rise time and the oscillation amplitude during burst tail is -0.02 with a probability of obtaining such correlation by a random data sample (i.e., chance probability) as 88.51\%. This supports the conclusion by \citet{Mahmoodifar2016} regarding the mechanism of burst decay oscillations. 

The presence and detection of burst oscillation can be triggered by a number of physical factors. One key factor that may contribute to the oscillation properties is the intensity. We then investigated the dependence of the oscillation amplitude on the average intensity during the oscillation phase. We present in Fig.~\ref{fig4} the variation of fractional amplitude with intensity for three individual sources. As evident from the plots, the amplitude shows a weak anti-correlation with the average intensity. The values of the correlation coefficient for 4U 1636--536, 4U 1702--42 and 4U 1728-34 are -0.44, -0.66 and -0.70 with chance probabilities of $5.7\times10^{-2}$, $7.8\times10^{-3}$ and $3.6\times10^{-2}$ respectively. Note that the peak intensity of the burst carries information about the ignition conditions, the burning fuel and the radiative processes during the burst. Therefore, we also examined the relation between the peak intensity and the oscillation amplitude. The behavior between these two quantities is displayed in Fig.~\ref{fig5} again for the same three classic sources. In this case, any correlation was weak; The correlation coefficient and chance probability values were obtained to be -0.37 and $1.2\times10^{-1}$ for 4U 1636--536, -0.45 and $8.9\times10^{-2}$ for 4U 1702--42, and -0.50 and $1.7\times10^{-1}$ for 4U 1728--34. 

The profile of a burst is expected to vary significantly depending on the latitude of the ignition site on the compact star. It is indeed demonstrated that the shape of the rise profile is correlated with the latitude at which the thermonuclear burning initiates \citep{Maurer2008}. The convexity parameter is used to quantify the variability of the shape of burst rise profile. The bursts that ignite near the poles show both positive and negative convexity whereas those which ignite near the equator always correspond to positive convexity values. Motivated by this distinction, we proceeded on to calculate the convexity parameter to infer the ignition latitude of the bursts and its possible correlation with the other burst timing and spectral characteristics. Following \citet{Maurer2008}, to calculate this parameter we have normalized the burst rise light curve so that the normalized burst rise count rate and burst rise time axes runs from 0 to 10. The convexity parameter is defined as 
\begin{equation}
\mathcal{C} = \sum_{i=0}^{N} ( c_i - x_i )\Delta t
\end{equation}
where $c_i$ is the normalized count rate in each bin, $x_i$ is the expected count rate for a slope of unity, and $\Delta t$ is the time bin size. This quantity basically computes the excess integrated area of the curve that deviates from the unity linear slope. Though the bursts with oscillations show primarily positive convexity, negative convexity is also observed (Table.~\ref{bparspec}). Our results imply that ignition gets initiated near the poles for in a large fraction of the bursts showing burst decay oscillations.

We investigated the time evolution of the oscillation during the burst in order to understand how the developed asymmetries propagate throughout the event. For this purpose, we divided the burst duration into finer intervals of 0.5 s, in each such segment the phase folded light curve was generated using the frequency model computed earlier. In each segment, the profile was then modeled with a sinusoidal model as before and the fractional amplitude was computed. Along with this, the profile was also modeled using a constant model. An F-test analysis was carried out between the two models to determine the detection significance of the oscillation in each time bin: If the sine model was better than the constant one with at least $3 \sigma$ significance, then the oscillation in that particular segment was considered to be significant. The amplitudes during the burst tail in all the cases exhibit significantly high value at least a few seconds after the burst peak as displayed in Fig.~\ref{fig08}. Following this interval, the amplitude displays lower values and generally does not correspond to a significant detection. The duration during which the oscillation was found to be significant according to our criterion is pointed out using vertical dashed lines in Fig.~\ref{fig08}. In Table.~\ref{bparosct} we provide the details of the maximum amplitude thus obtained during the burst and its time of occurrence relative to the burst peak. As it can be seen from this table, for oscillation during the tail the oscillation reaches peak power a few seconds after the burst peak. The negative times in this table pertains to the oscillations detected during burst rise only.

Along with temporal aspects, we also studied spectral evolution of thermonuclear bursts to understand the mechanism behind the evolving properties of burst oscillations. We selected the burst interval starting from the rise time to the time where the intensity drops down to 10\% of the burst peak intensity. For performing the time-resolved spectroscopy we divided the burst interval into finer segments of 1 s duration. For each segment, we extracted the burst spectrum using event mode data. The background spectrum was extracted from a 10 s long persistent (non-burst) emission interval of the light curve. The response files were then generated using the standard FTOOLS recipe PCARSP considering all active PCUs. Burst spectra are reasonably well fitted with a blackbody model as shown in previous literature \citep{Galloway2008}. Therefore, we fitted each spectrum using XSPEC version 12.9.0n and with an absorbed blackbody model (i.e., {\tt tbabs*bbodyrad}). The neutral hydrogen column density for each source was kept constant at the value obtained from high resolution spectral observations reported in the literature. The evolving blackbody temperature and normalization were noted for each spectrum and the corresponding flux was computed using the \textit{cflux} tool. From the time evolving flux values, the total fluence of the burst was computed by integrating the flux over the entire burst duration. 

As expected, the blackbody model provided an adequate representation of most of time resolved X-ray spectra. The fit quality was poor in some cases most likely due to relatively large integration time of 1 s. Note also the fact that we have not included any systematic uncertainty in our fits. We list the averaged blackbody temperatures in Table 3. We find that the blackbody temperature varied between 1 and 3 keV. A fraction of the more luminous bursts exhibited photospheric radius expansion phase (PRE) displaying a temperature minimum and normalization maximum near the peak of the burst. The temperature follows a steady decaying trend as it is a characteristic of thermonuclear bursts, while the normalization usually shows an increasing trend in these bursts possessing burst oscillations. In Fig.~\ref{fig6} we present an example of the spectral evolution during such a burst. 

We investigated the dependence of the oscillation amplitude on the simultaneous blackbody temperature to seek for any connection between the two parameters. We present in Fig.~\ref{fig7}, the variation of fractional amplitude with respect to the maximum blackbody temperature within the corresponding oscillation interval. The red data points represent those temperatures where the oscillation was detected during the rising phase of the burst. We find no connection between the maximum blackbody temperature during the duration of significant oscillation and the amplitude of oscillation (the correlation coefficient is -0.27 with a chance probability of $5.3 \times 10^{-2}$). Since the blackbody temperature declines steadily during the burst tail, and if the oscillation duration is long enough, the blackbody temperature might have changed during the time frame of oscillation. For this reason, we also checked the average blackbody temperature during the oscillation duration and compared its effect on the oscillation amplitude. Again we find no correlation: the correlation coefficient is -0.42 with a chance probability of $2.4\times10^{-3}$.

\section{Discussion}\label{discussion}

Thermonuclear X-ray bursts with transient oscillations are unique tools to delve into the physics behind the exotic environments of extreme densities. Systematic studies of these events to uncover their temporal and spectral characteristics have been ongoing. Our investigation here is the latest addition to the sequence of extensive studies with joint timing and spectral characterization of burst oscillations. Such systematic investigations are essential for further studies with currently operating X-ray telescopes like Astrosat LAXPC \citep{Agrawal2006} and NICER \citep{Gendreau2012}, as well as future missions, such as eXTP \citep{snZhang2016}.

In this paper we have extensively investigated the phase and amplitude evolution of burst oscillation with energy. We studied the frequency variation of oscillation during the thermonuclear bursts and observed that the frequency drifts are more frequent in the cases where the spin frequency is lower \citep{Muno2002}. We found that the phase lag of the burst oscillations shows no systematic evolution with energy between the bursts and also between the different sources. Though in a number of cases we do indeed observe lag of the soft energy photons, there are a significant number of cases for which hard lag or no lag is observed. Unlike \citet{Muno2003}, we have not averaged the profiles of oscillations from different bursts, since the properties of bursts, such as, ignition latitude, burning fuel may vary widely between different bursts even from the same source. This was an issue raised by \citet{Artigue2013} about the methodology of \citet{Muno2003}. We found that the amplitude of the burst oscillations, on the other hand, primarily showed a monotonic increase with energy, as it is in accordance with the theoretical expectations. The amplitude of oscillations is expected to be lower in the softer energy bands as continuum emission from the neutron star peaks in the lower energy bands decreasing the brightness contrast. The evolution of the amplitude with energy varied between bursts even from the same source again, possibly due to different thermonuclear ignition conditions. The amplitude evolution depends on the brightness contrast which is a function of the temperature distribution, the ignition latitude, the fuel composition, the hot spot size and shape. Between bursts from different sources, the stellar spin and the redshift due to the stellar gravity (a function of M/R) could also affect the oscillation amplitude behavior. Here we note that for accretion powered pulsations, the amplitude occasionally shows a decreasing trend with energy contrary to the burst oscillations \citep{Muno2003, Watts2012}. 

\begin{figure*}
\centering
\begin{tabular}{ccc}
\textbf{4U 1636--536} & \textbf{4U 1702--42} & \textbf{4U 1728--34} \\
\noalign{\vskip 3mm}    
\includegraphics[width=0.25\textheight]{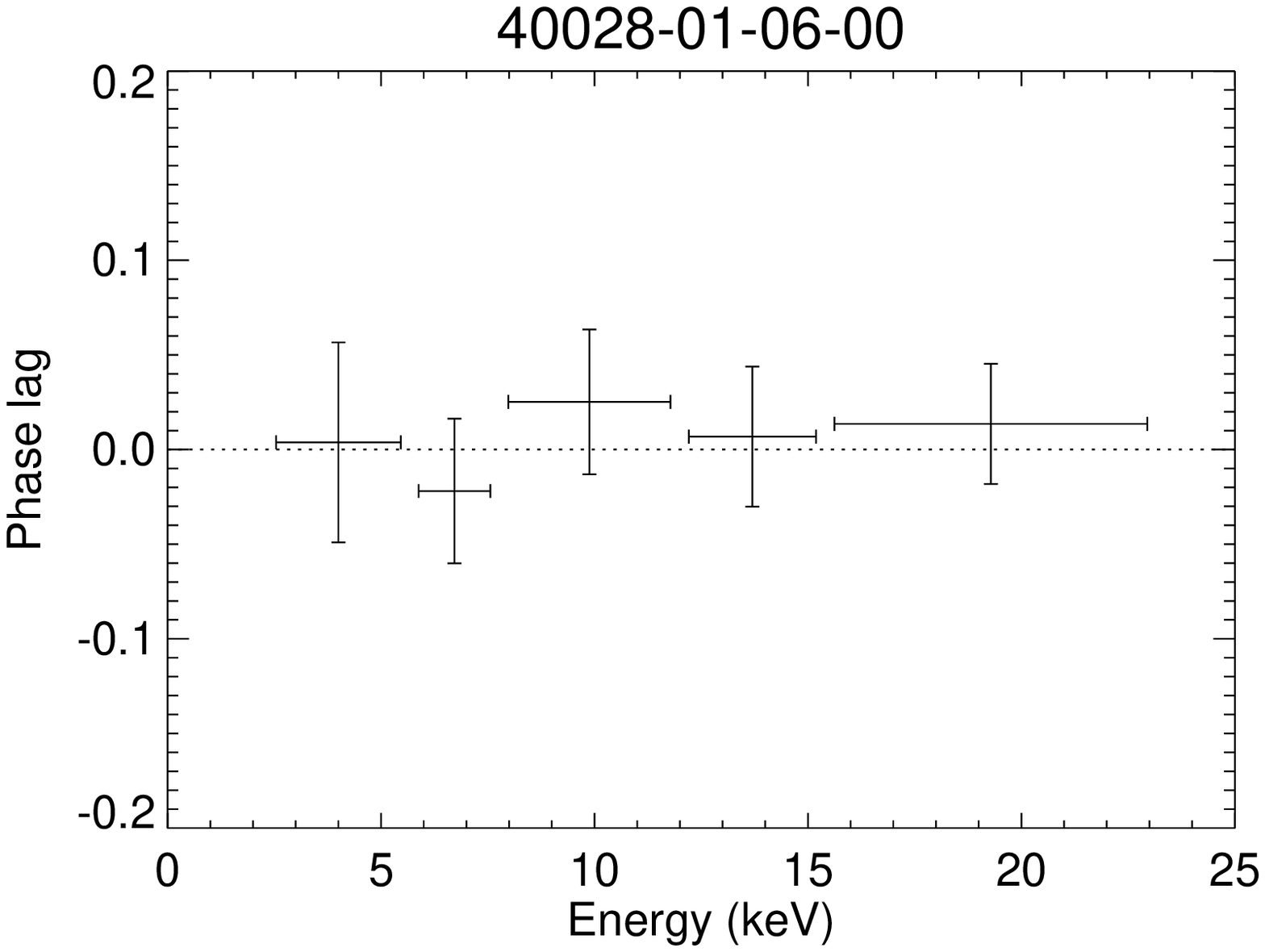} &
\includegraphics[width=0.25\textheight]{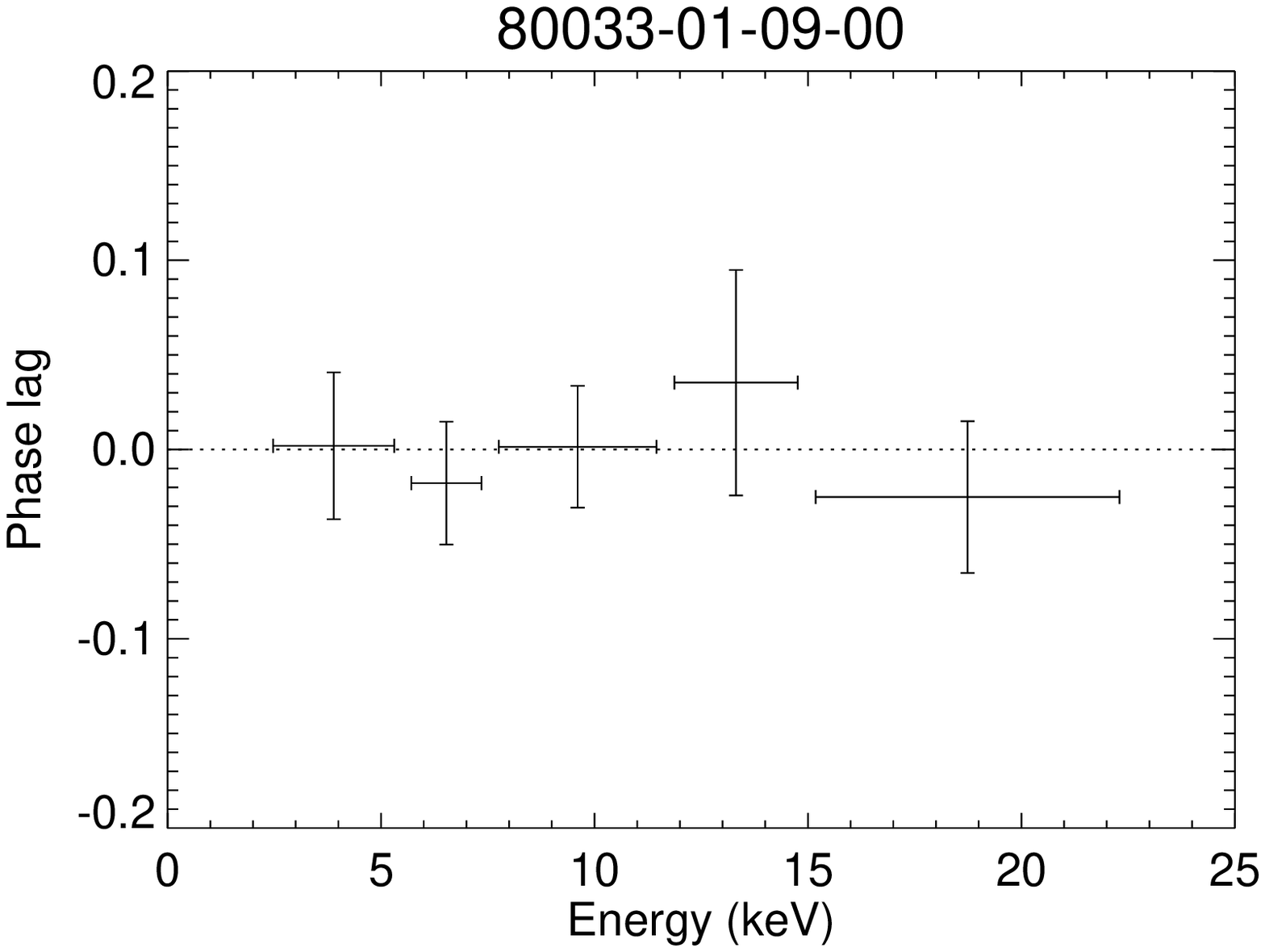} &
\includegraphics[width=0.25\textheight]{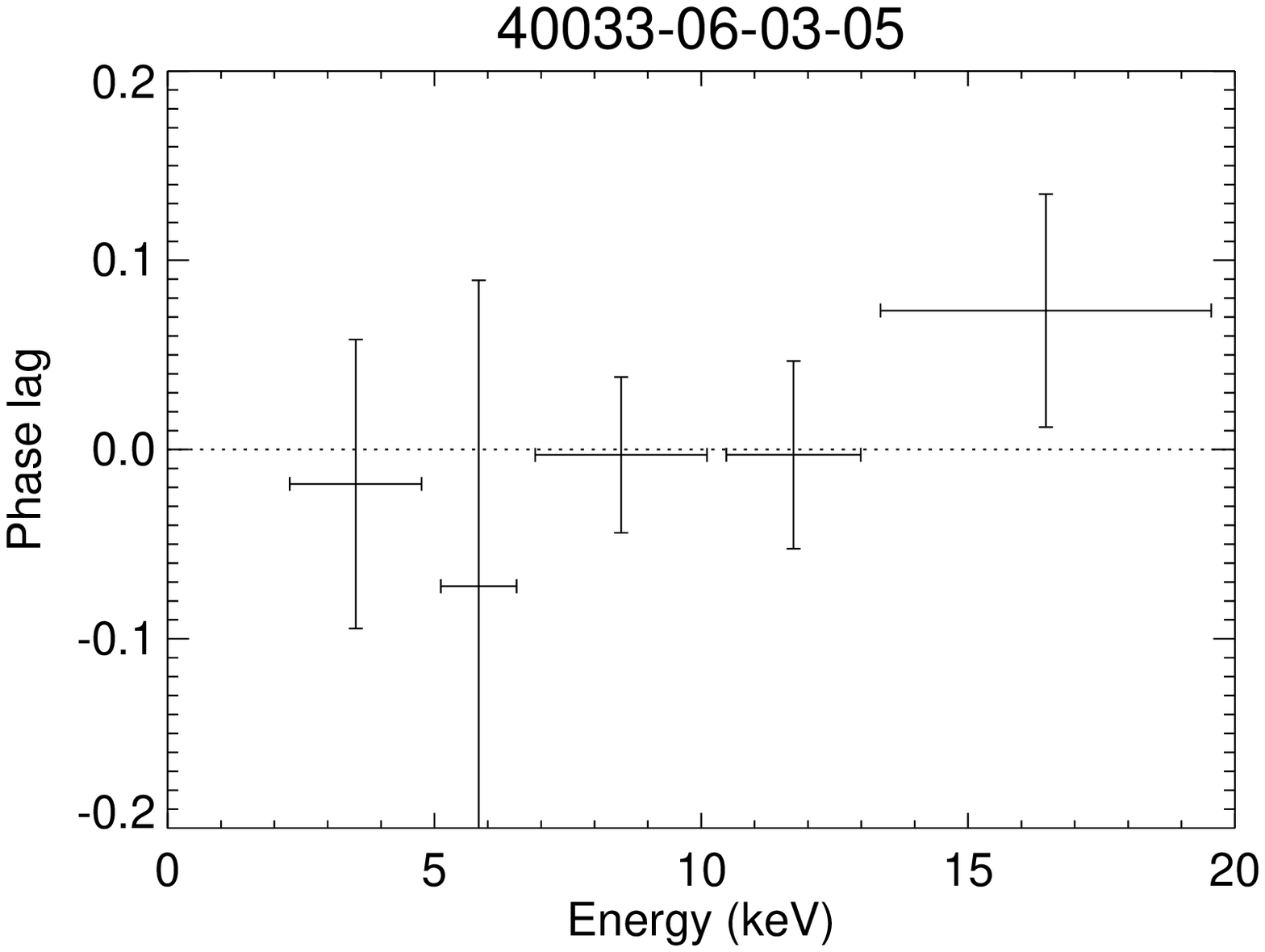} \\
\includegraphics[width=0.25\textheight]{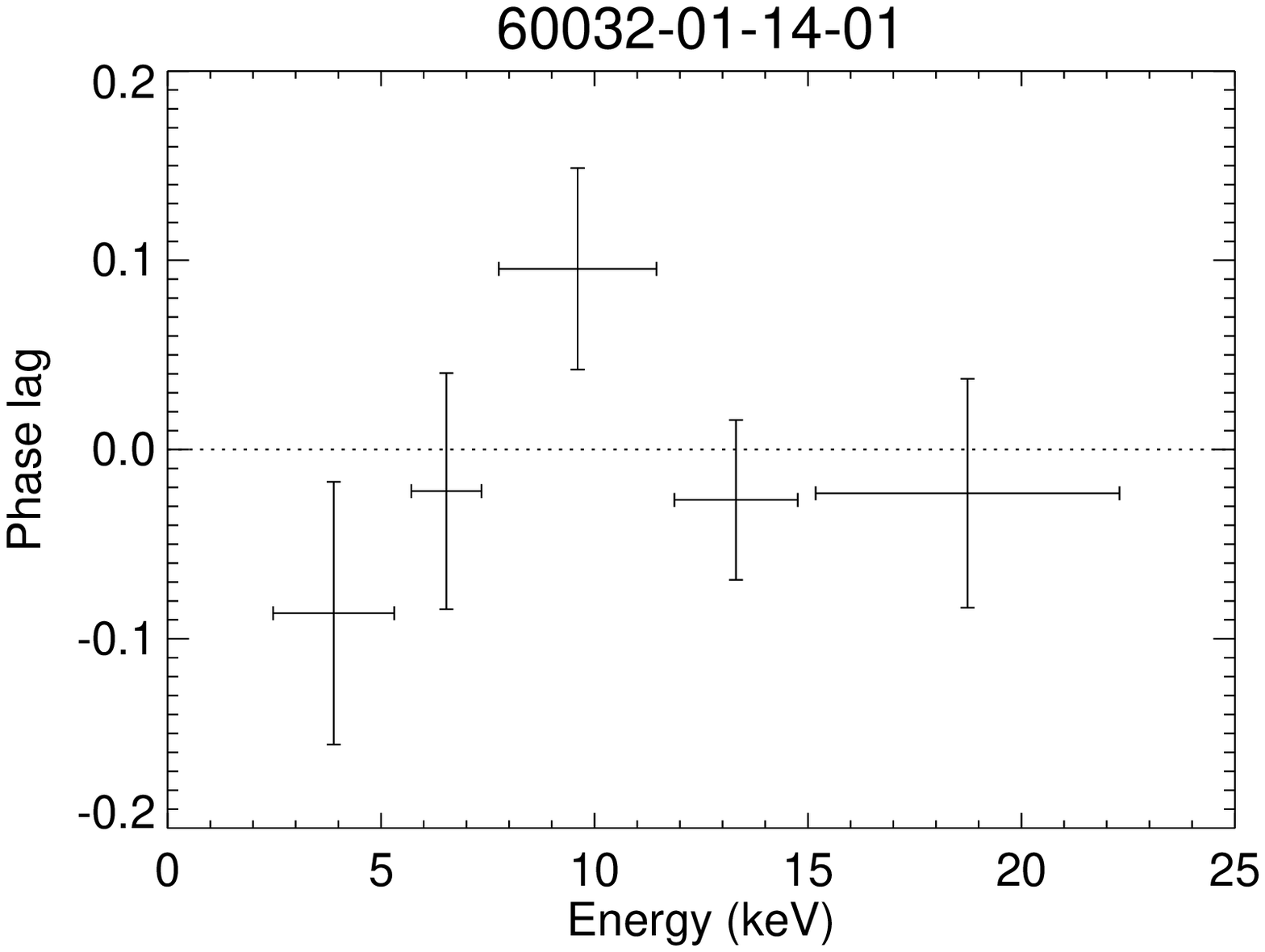} &
\includegraphics[width=0.25\textheight]{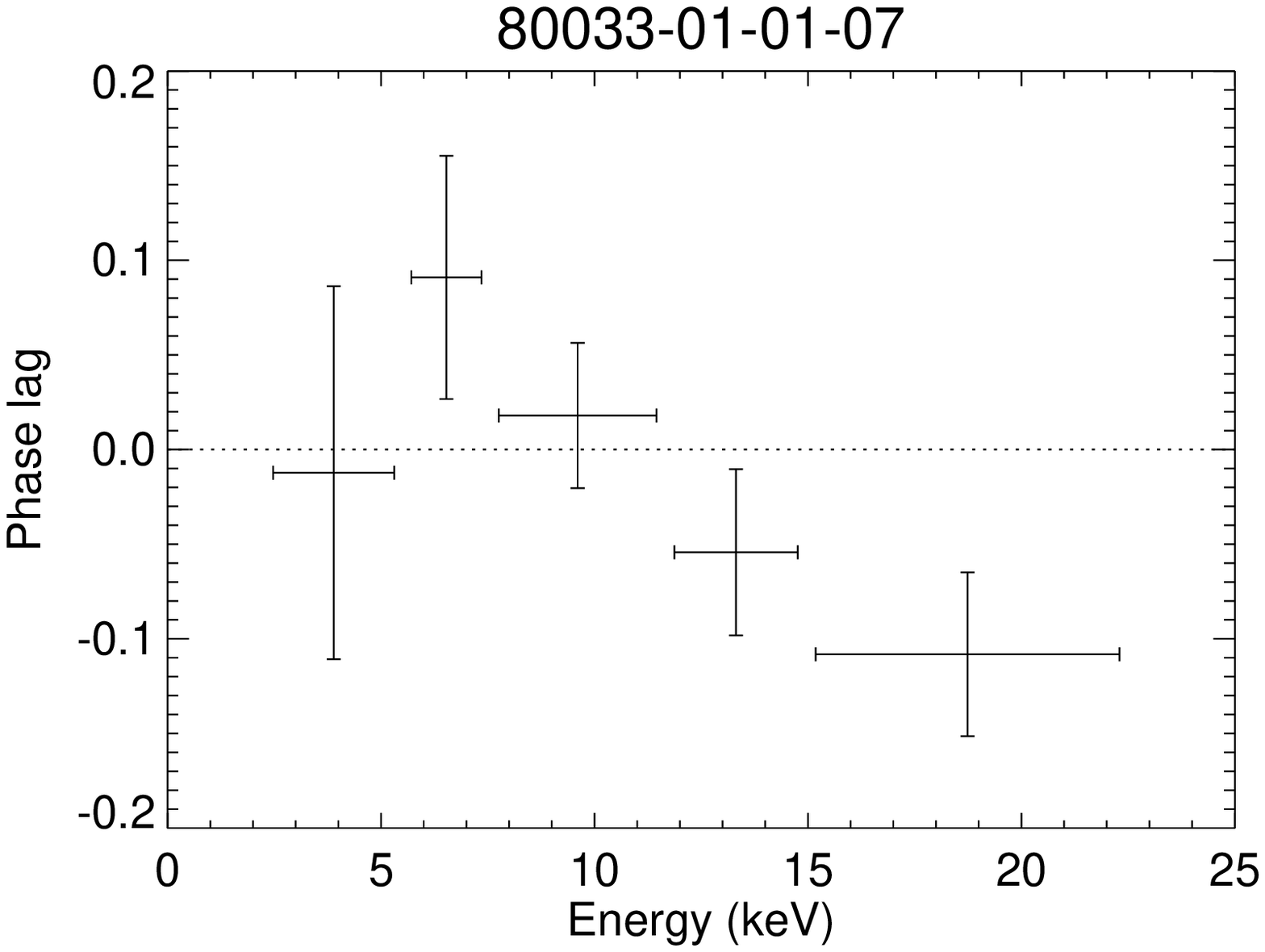} &
\includegraphics[width=0.25\textheight]{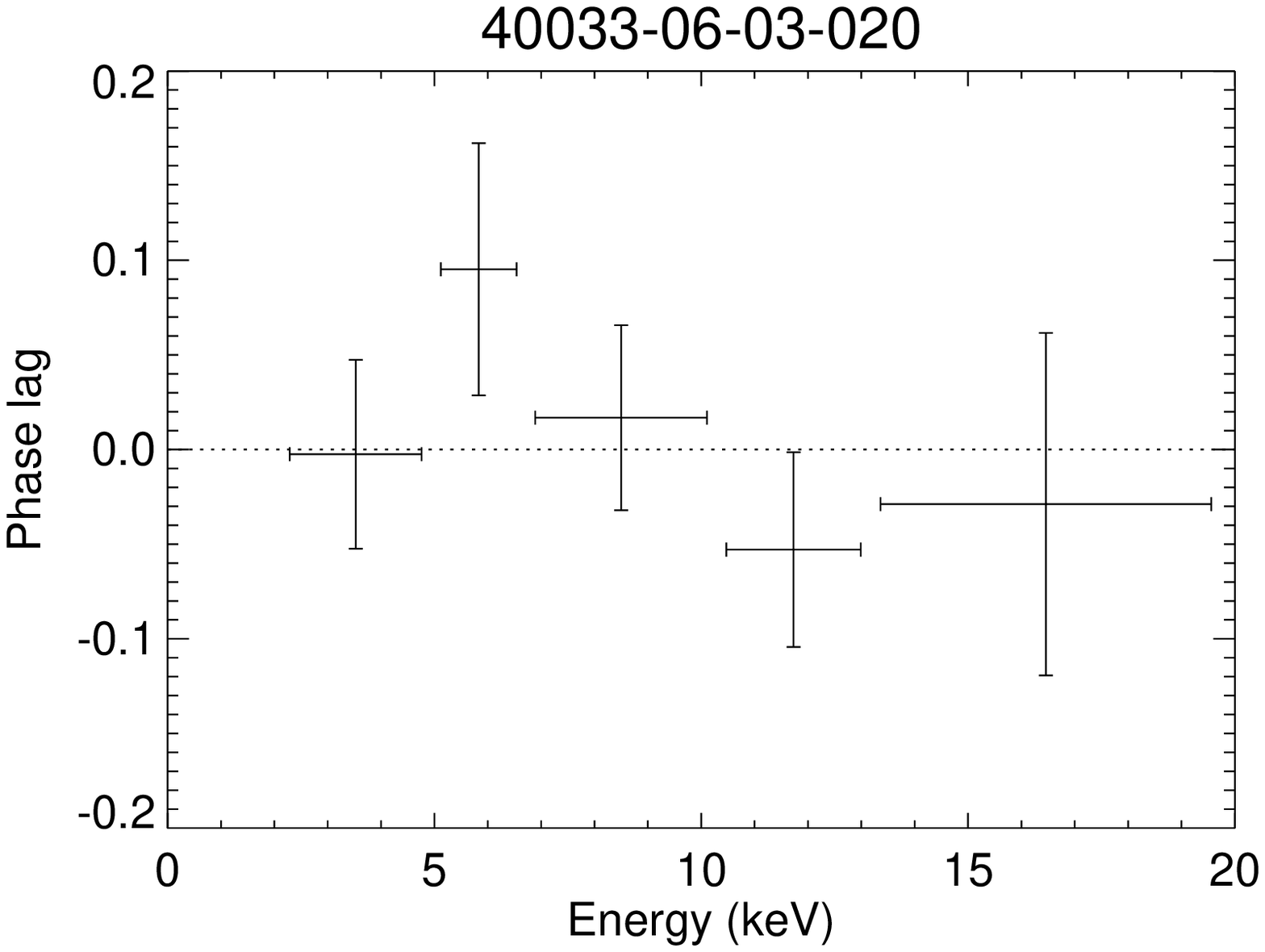} \\
\includegraphics[width=0.25\textheight]{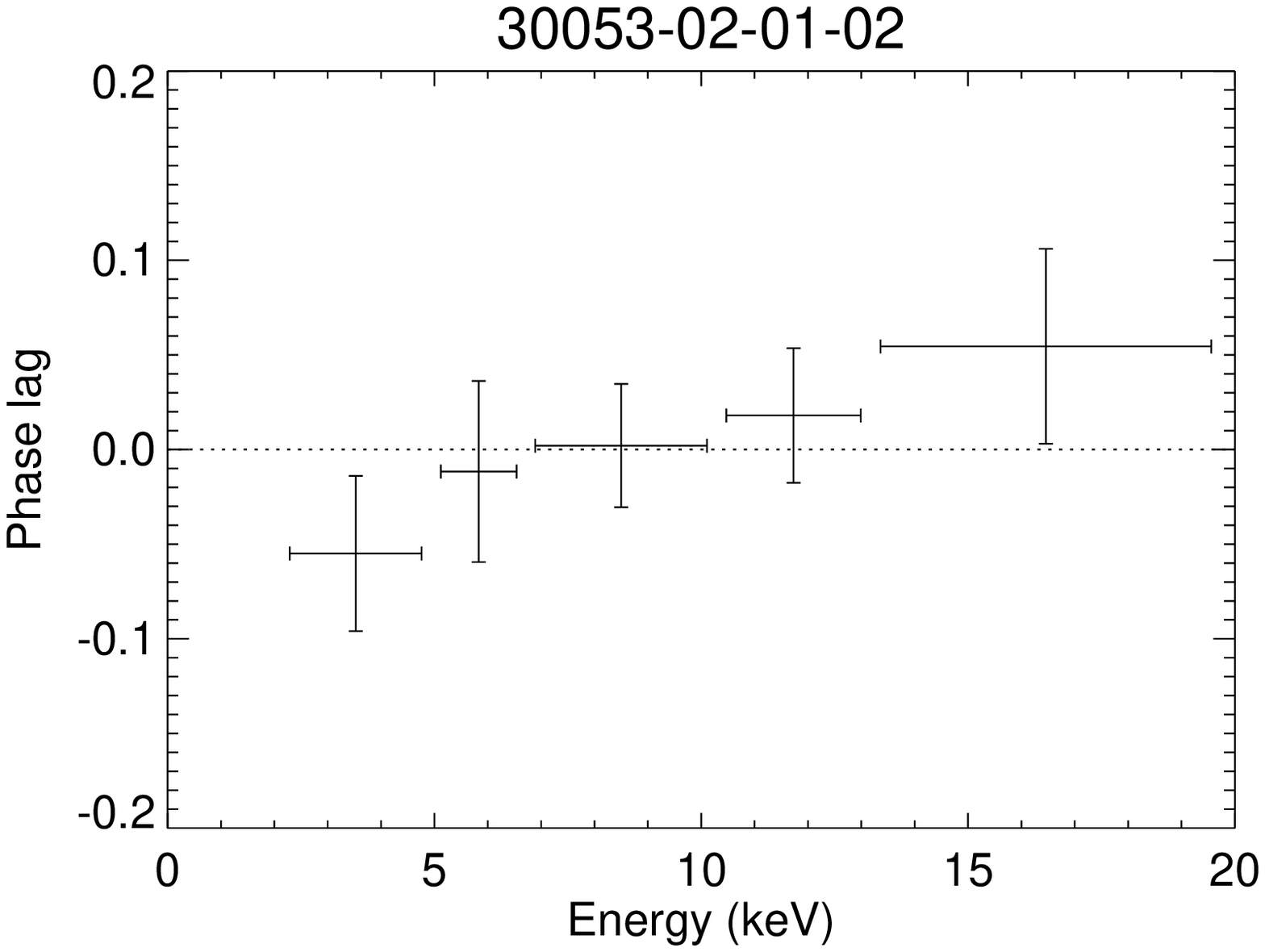} &
\includegraphics[width=0.25\textheight]{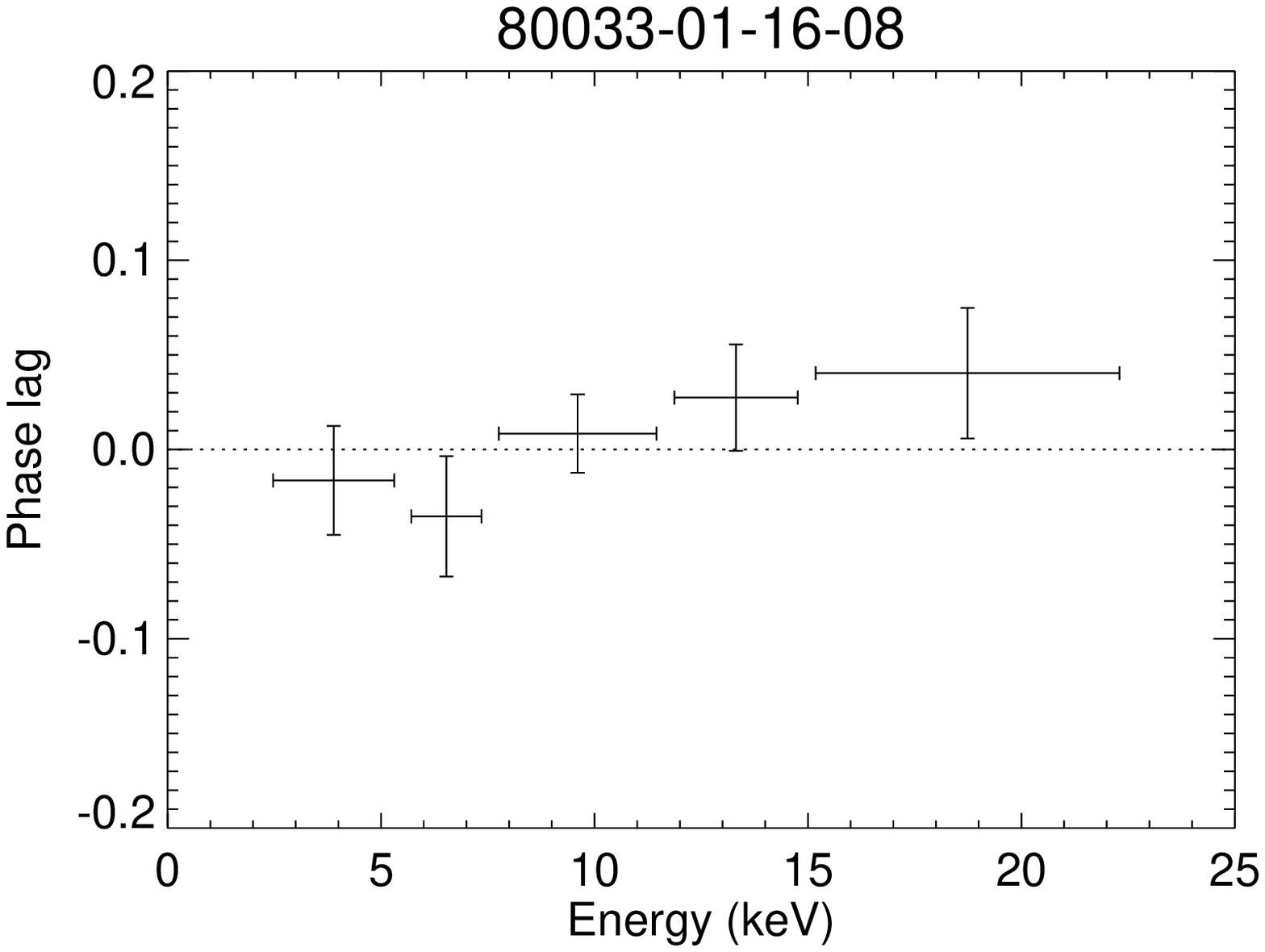} &
\includegraphics[width=0.25\textheight]{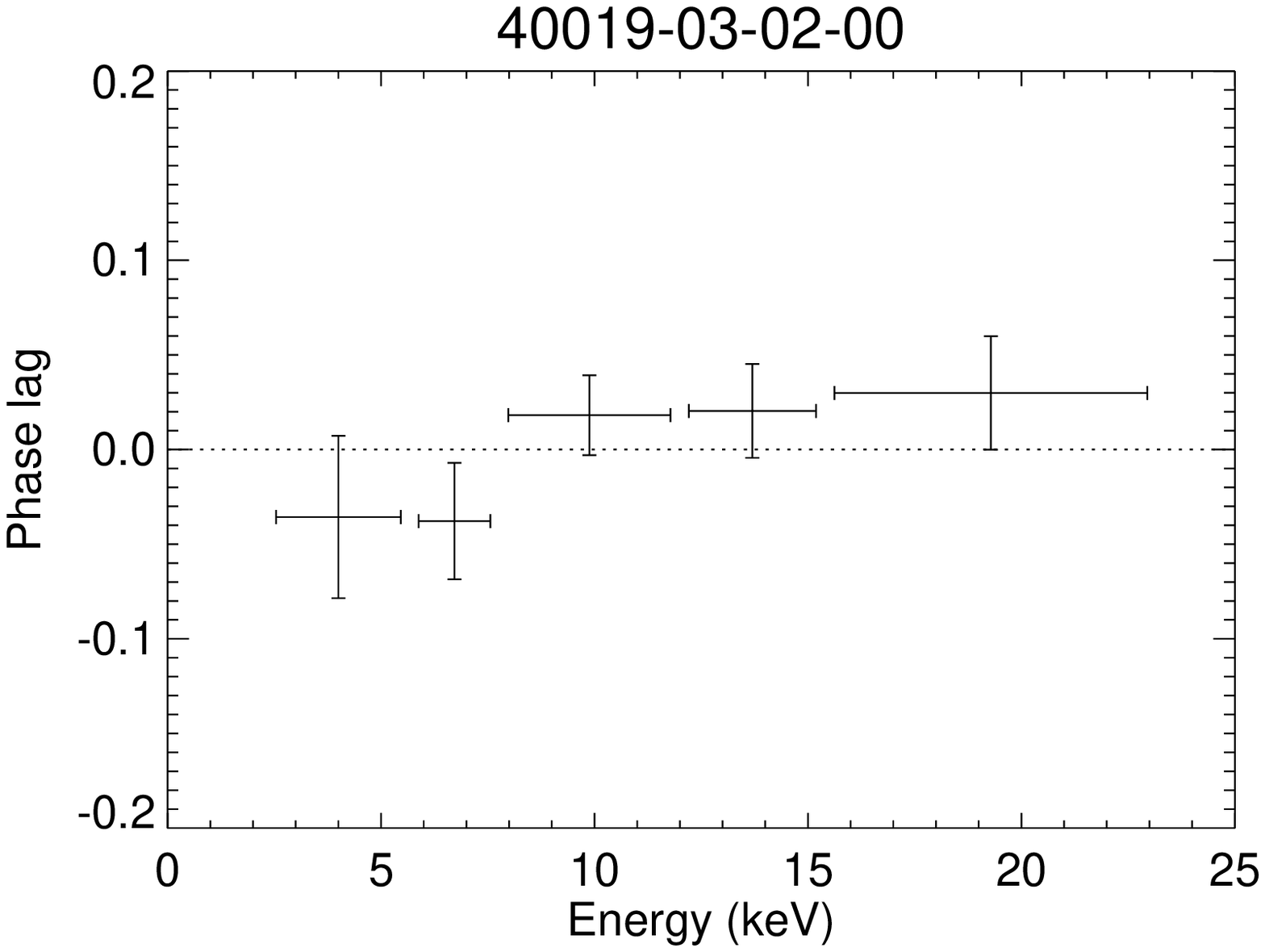} \\
\includegraphics[width=0.25\textheight]{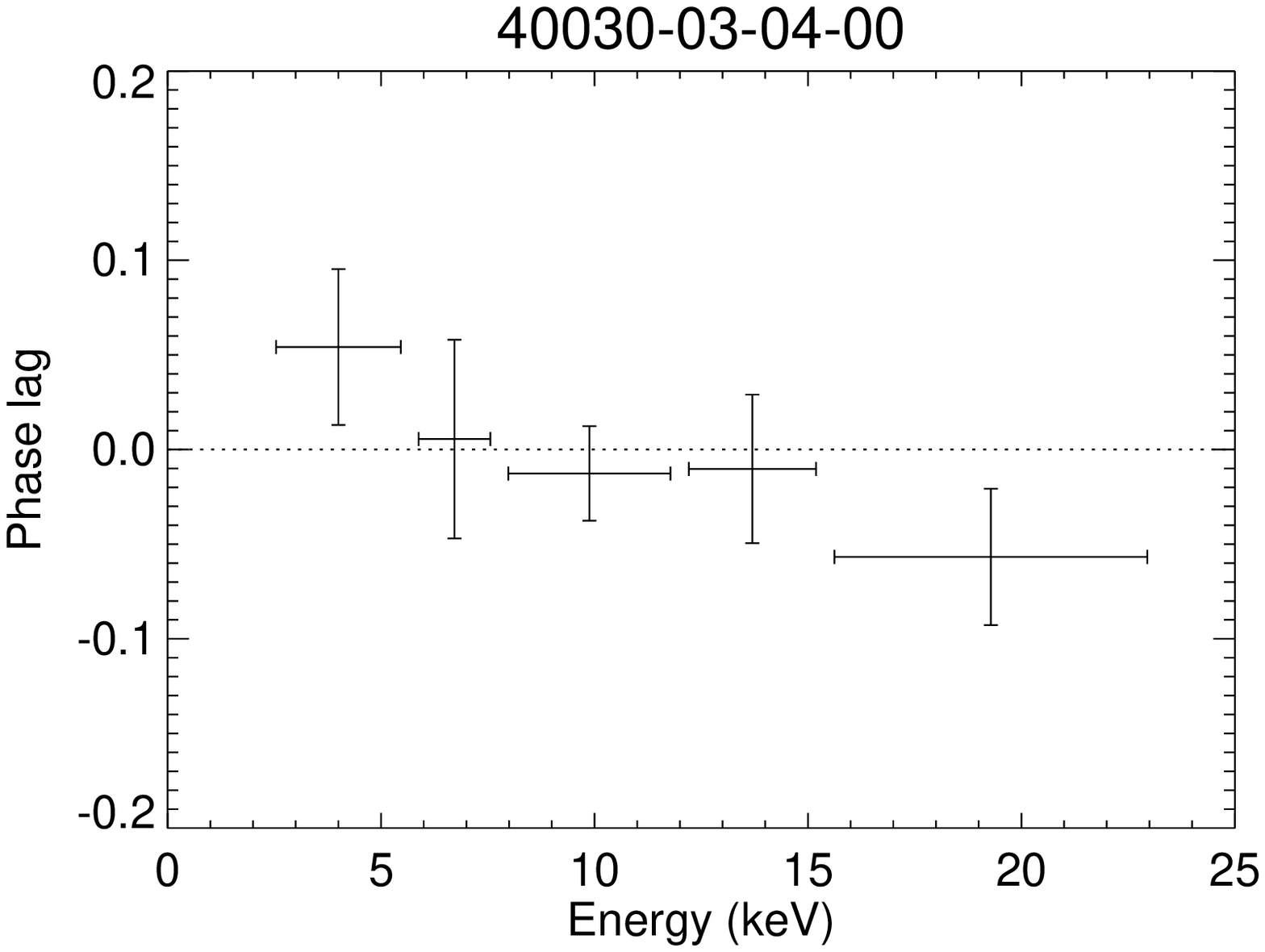} &
\includegraphics[width=0.25\textheight]{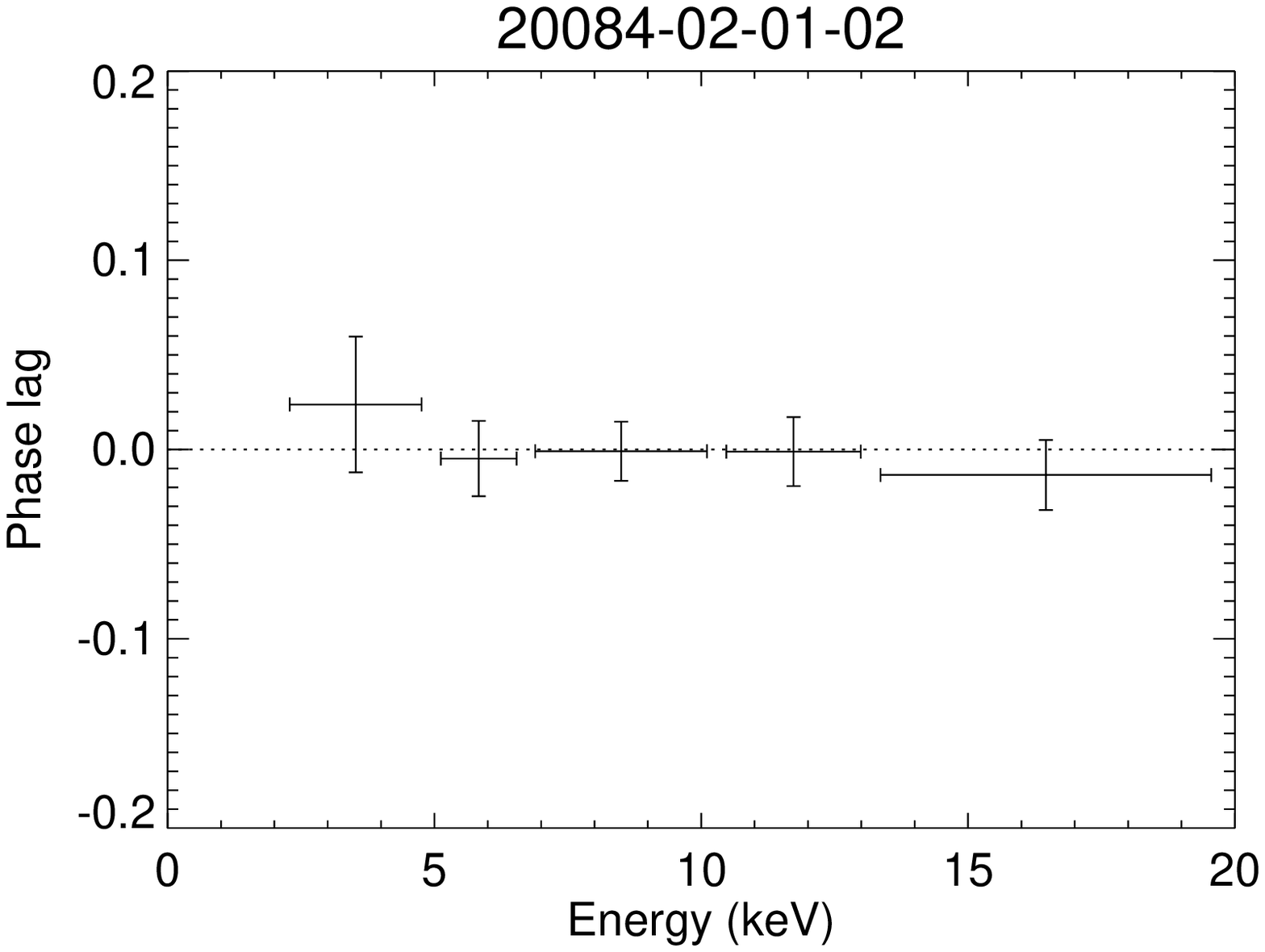} &
\includegraphics[width=0.25\textheight]{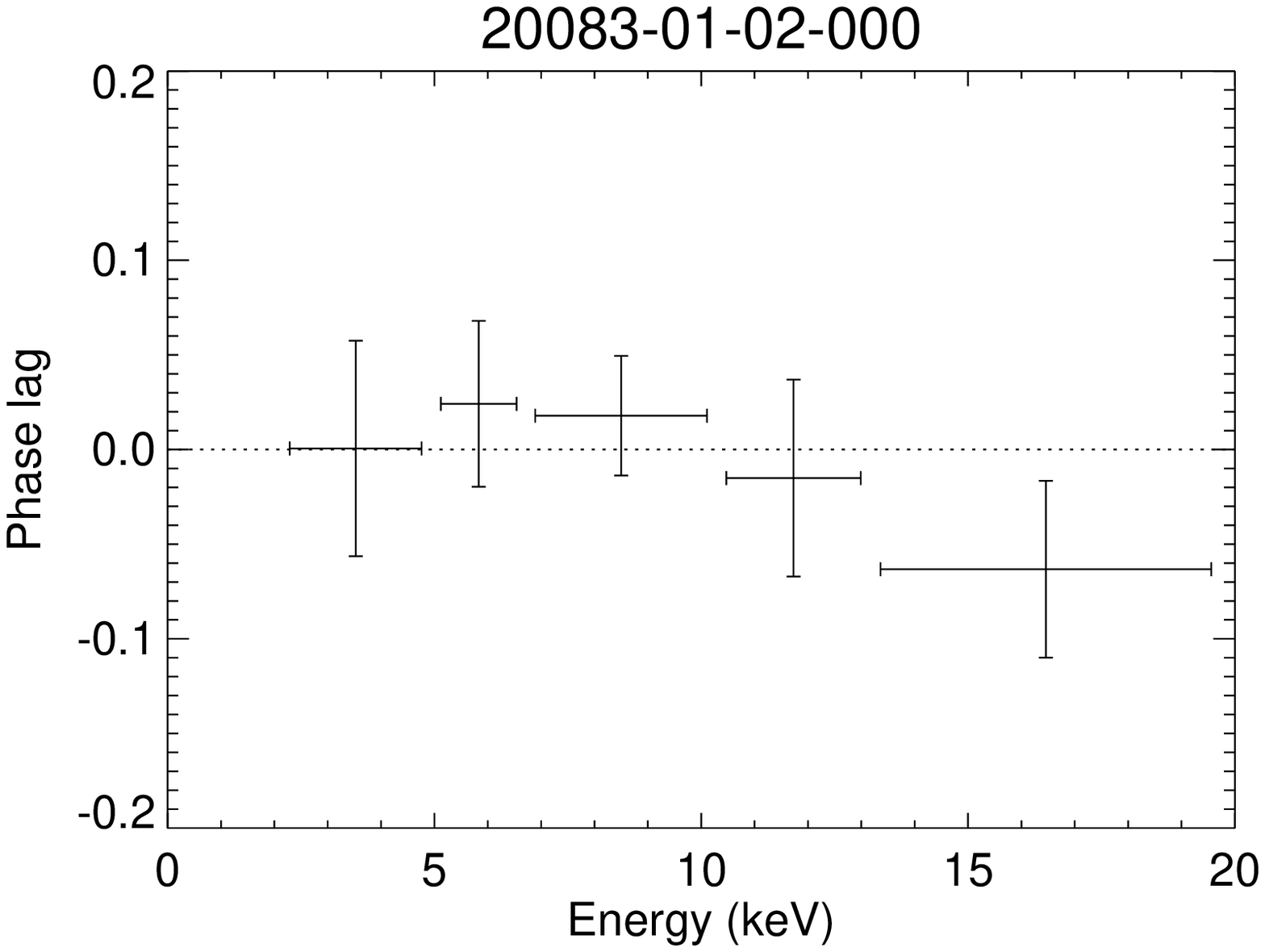} \\
\end{tabular}
\caption{Phase lag variation of burst oscillation with energy. The different columns represent the different sources and the different rows denote the different phase evolution with energy. Here we have chosen a representative sample to highlight the varying phase energy behavior across sources. The burst IDs considered for this figure are identified at the top of each panel. \label{fig01}}
\end{figure*}

\begin{figure*}
\centering
\begin{tabular}{cc}
\includegraphics[width=0.32\textheight]{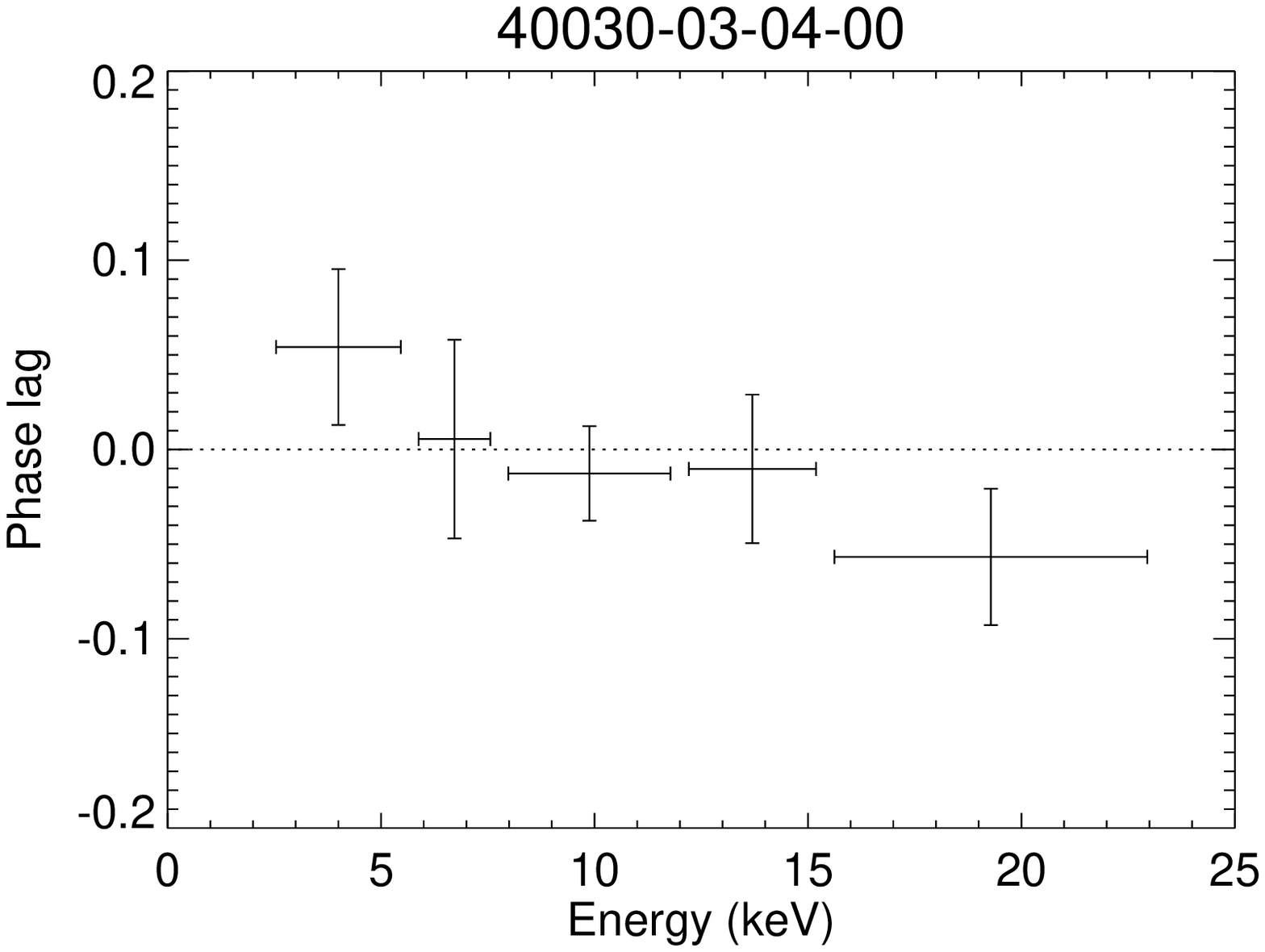} &
\includegraphics[width=0.32\textheight]{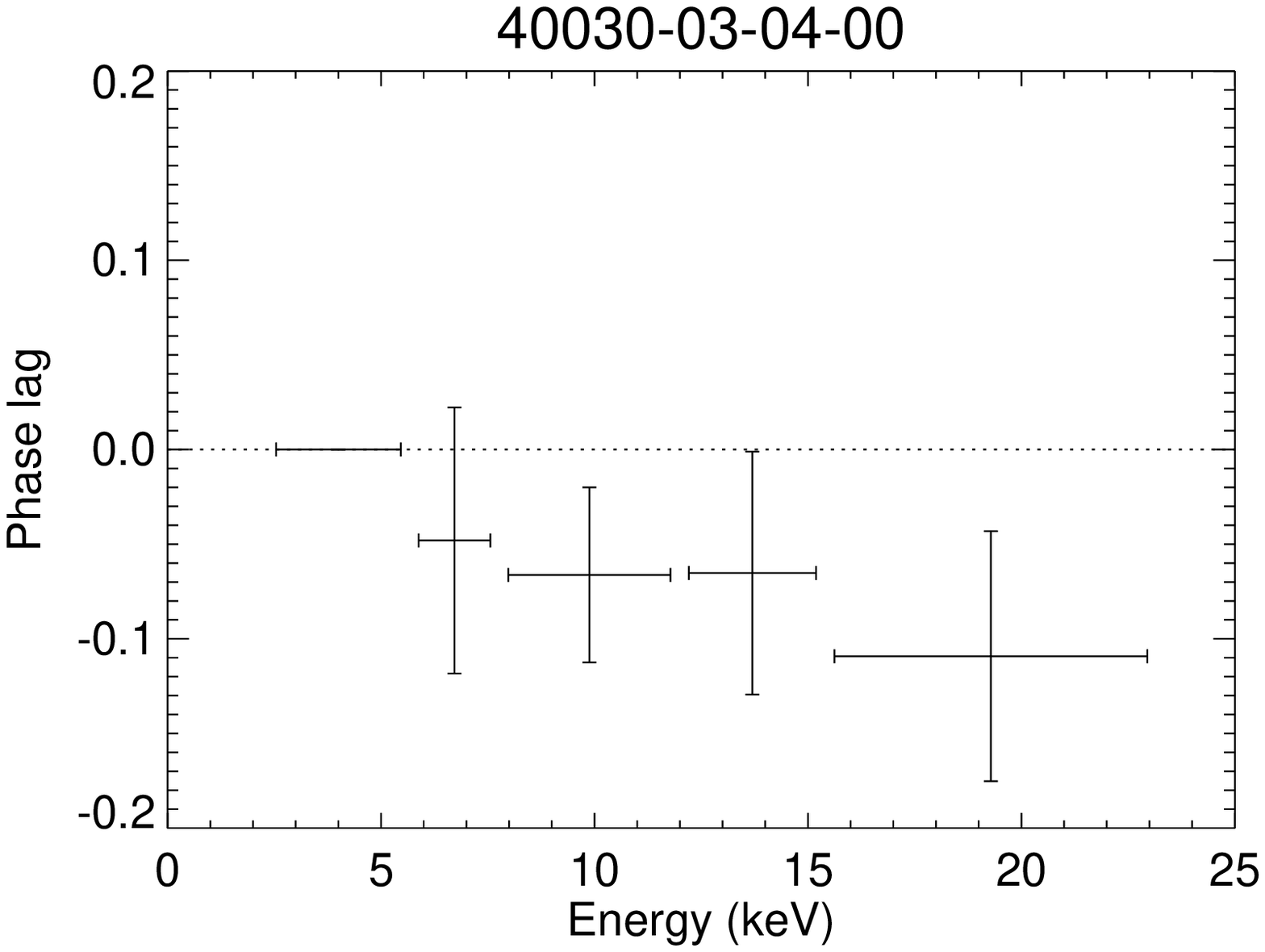} 
\end{tabular}
\caption{Comparison between the phase lag behavior for different reference band choice. On the left the phase lag evolution is computed considering the entire energy band and on the right panel the phase lag evolution is computed considering the first channel range (channels 5-12) as the reference band. In the later case no error are put on the first energy band as it is taken as the reference band. Between the two methods even though the magnitude of the phase lag changes as a consequence of changing the reference, the sign and the evolution exhibits the same behavior. The errors given here are $1 \sigma$ statistical errors on the phase difference and are computed propagating the errors on the phases obtained through sinusoidal fitting of the phase-folded lightcurves in each energy band. \label{fig1a}}
\end{figure*}

\begin{figure*}
\centering
\begin{tabular}{ccc}
\textbf{4U 1636--536} & \textbf{4U 1702--42} & \textbf{4U 1728--34} \\
\noalign{\vskip 3mm}    
\includegraphics[width=0.25\textheight]{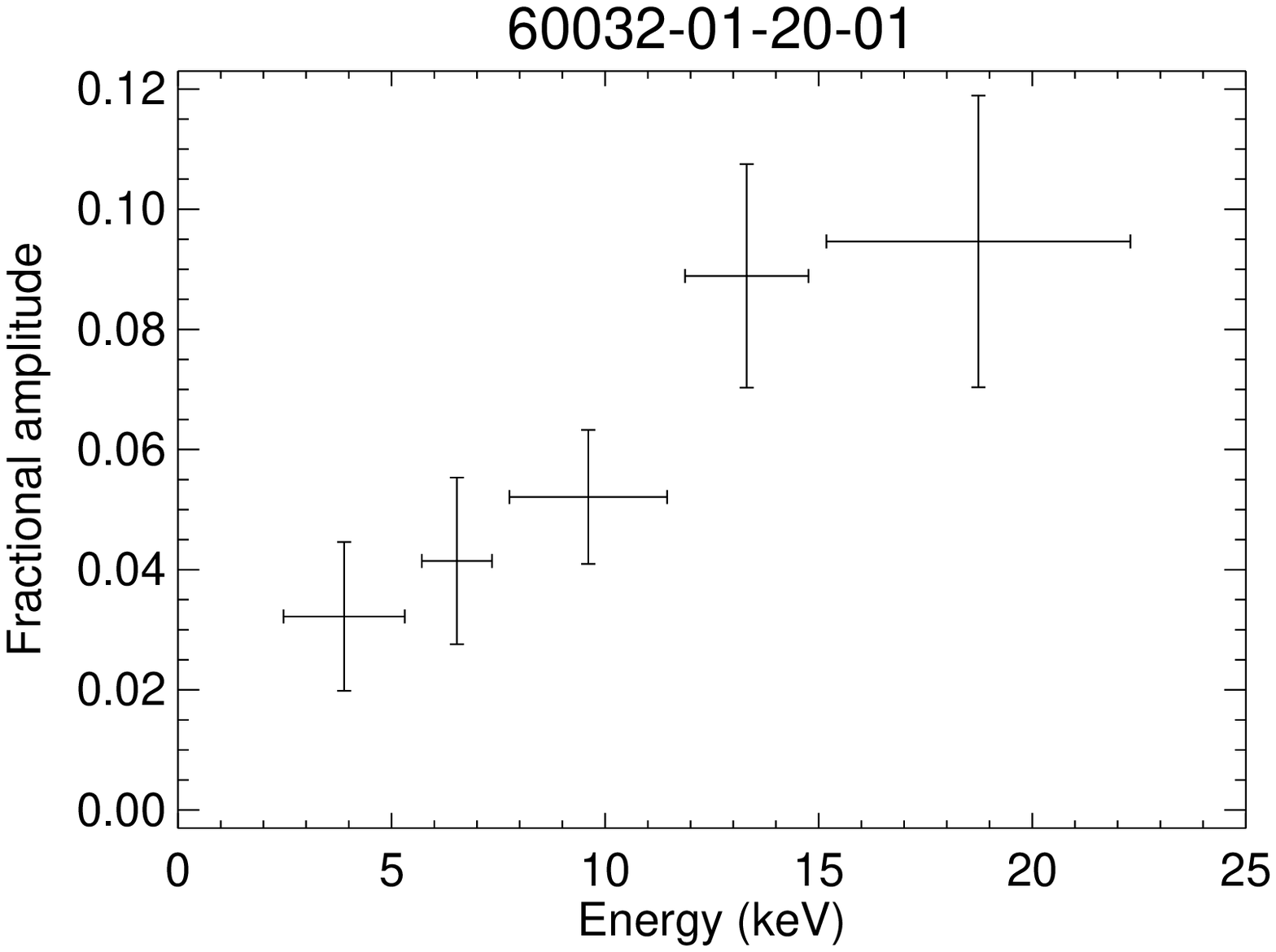} &
\includegraphics[width=0.25\textheight]{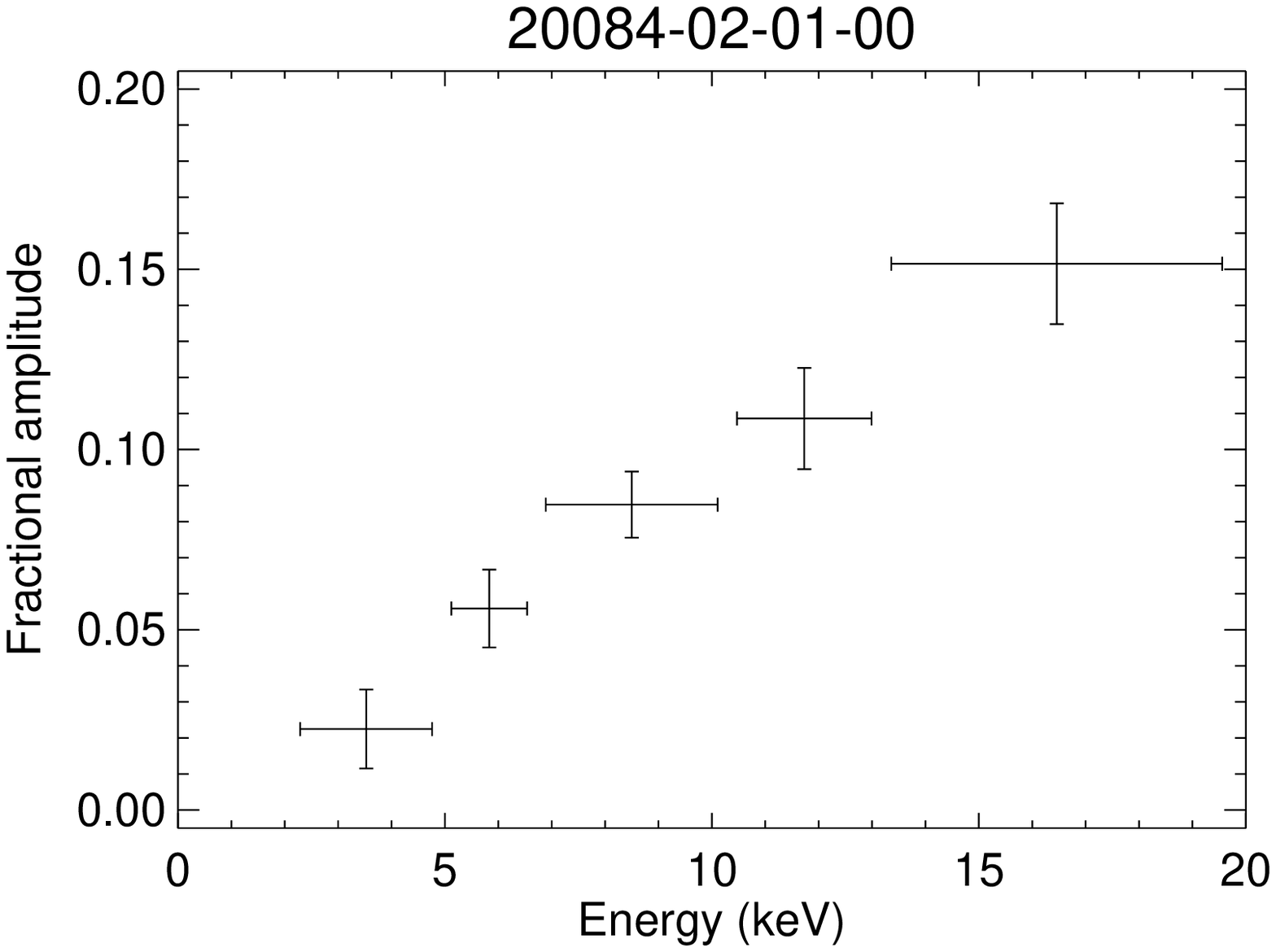} &
\includegraphics[width=0.25\textheight]{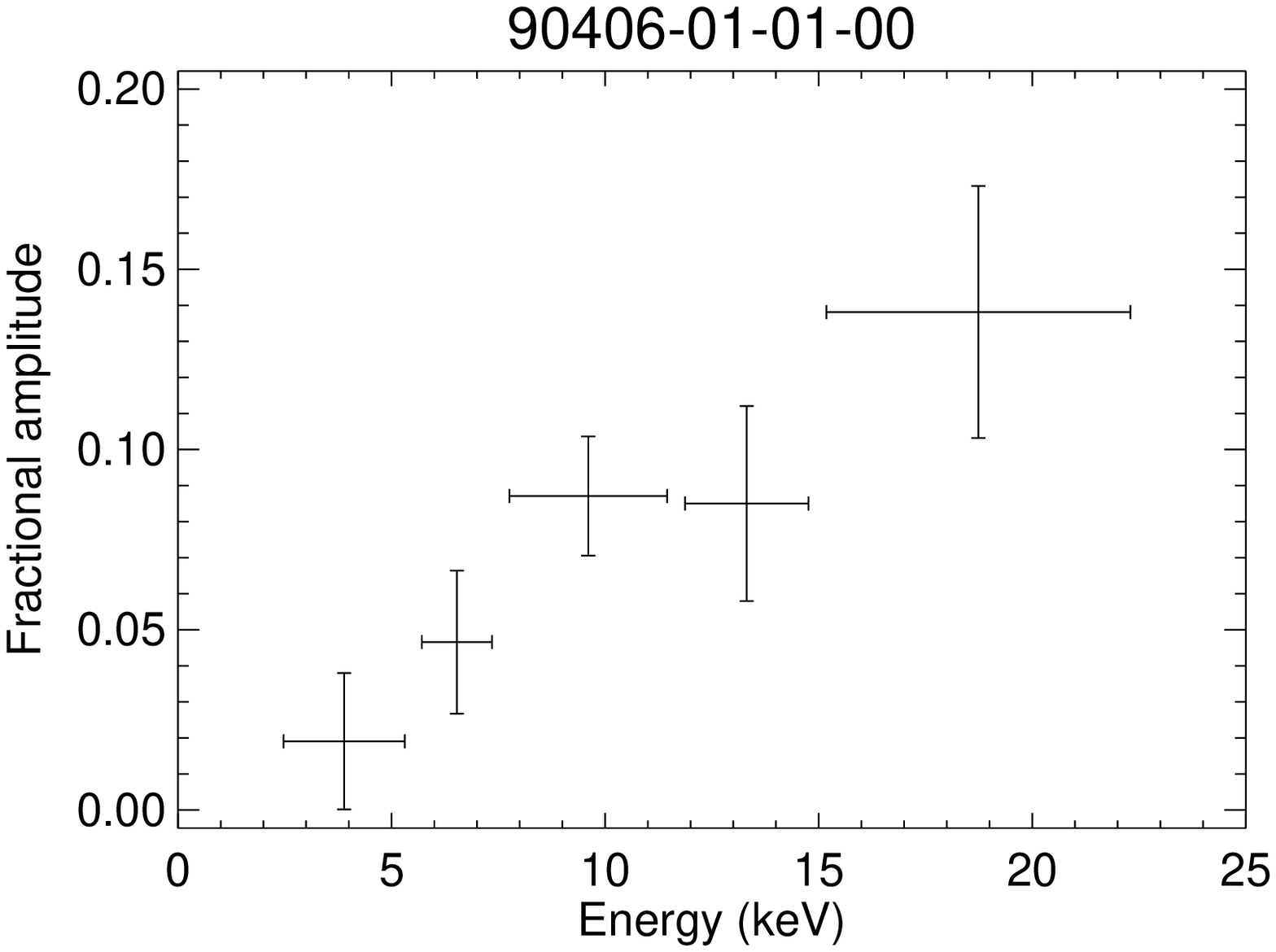} \\
\includegraphics[width=0.25\textheight]{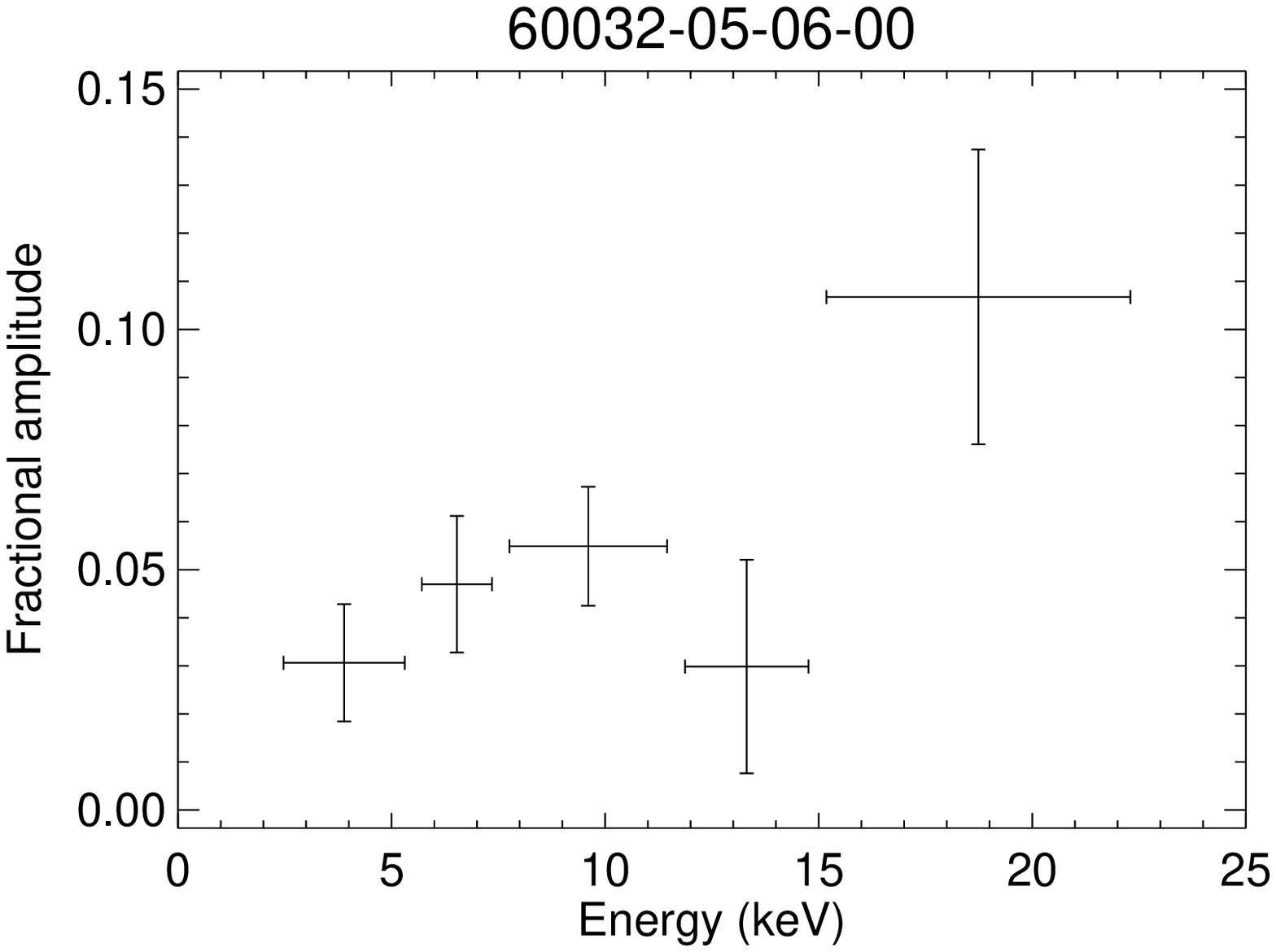} &
\includegraphics[width=0.25\textheight]{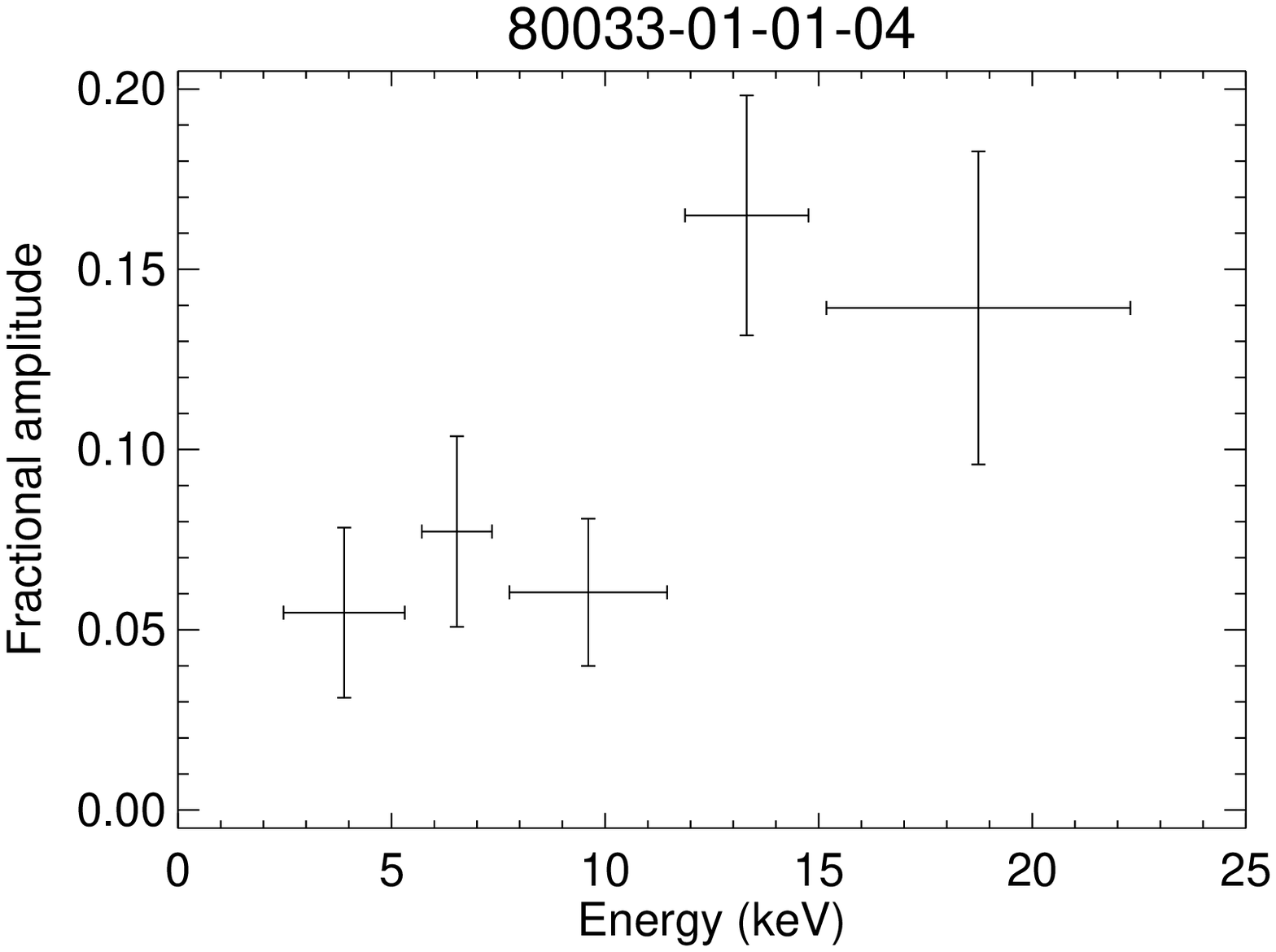} &
\includegraphics[width=0.25\textheight]{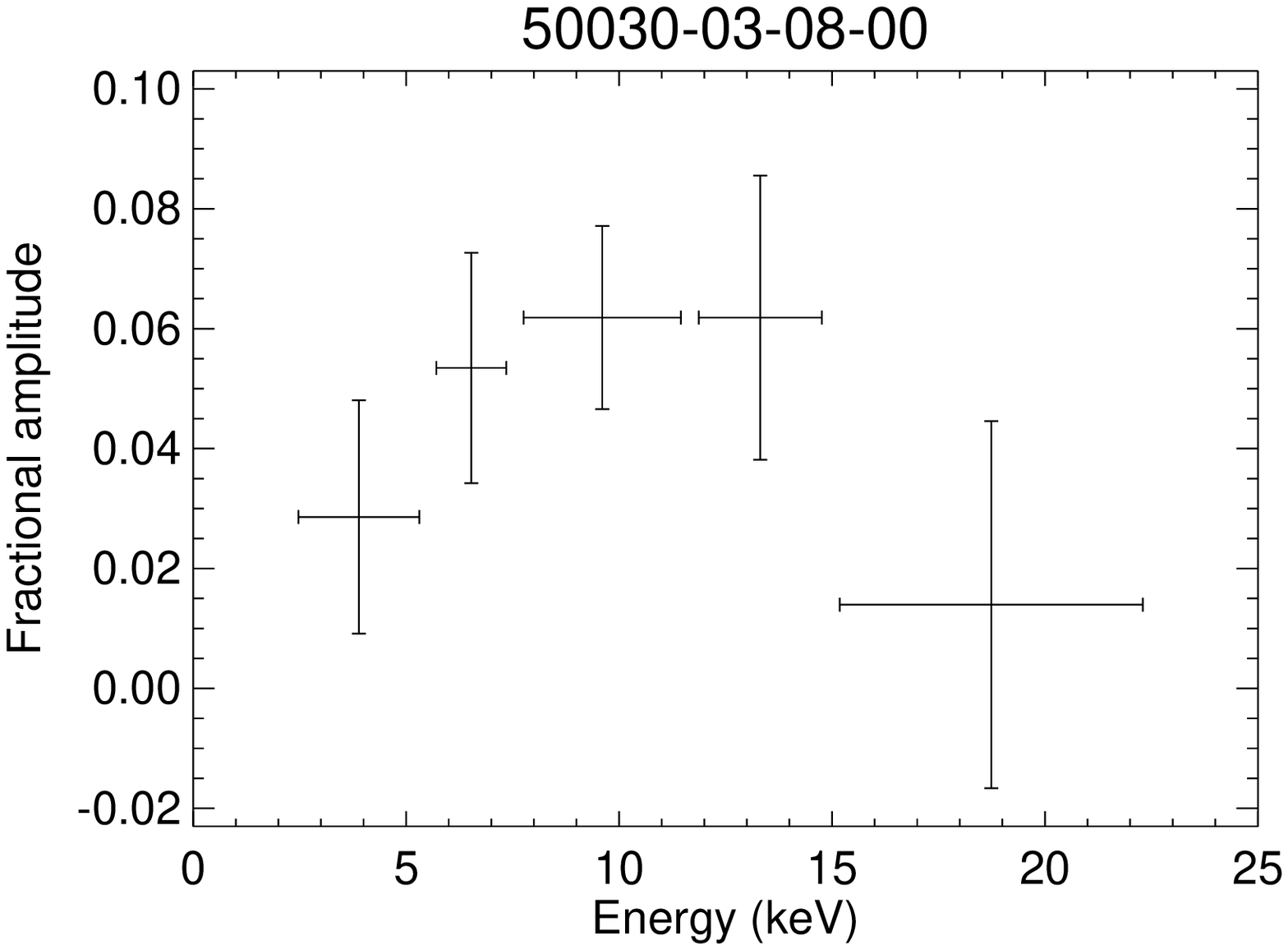} \\
\includegraphics[width=0.25\textheight]{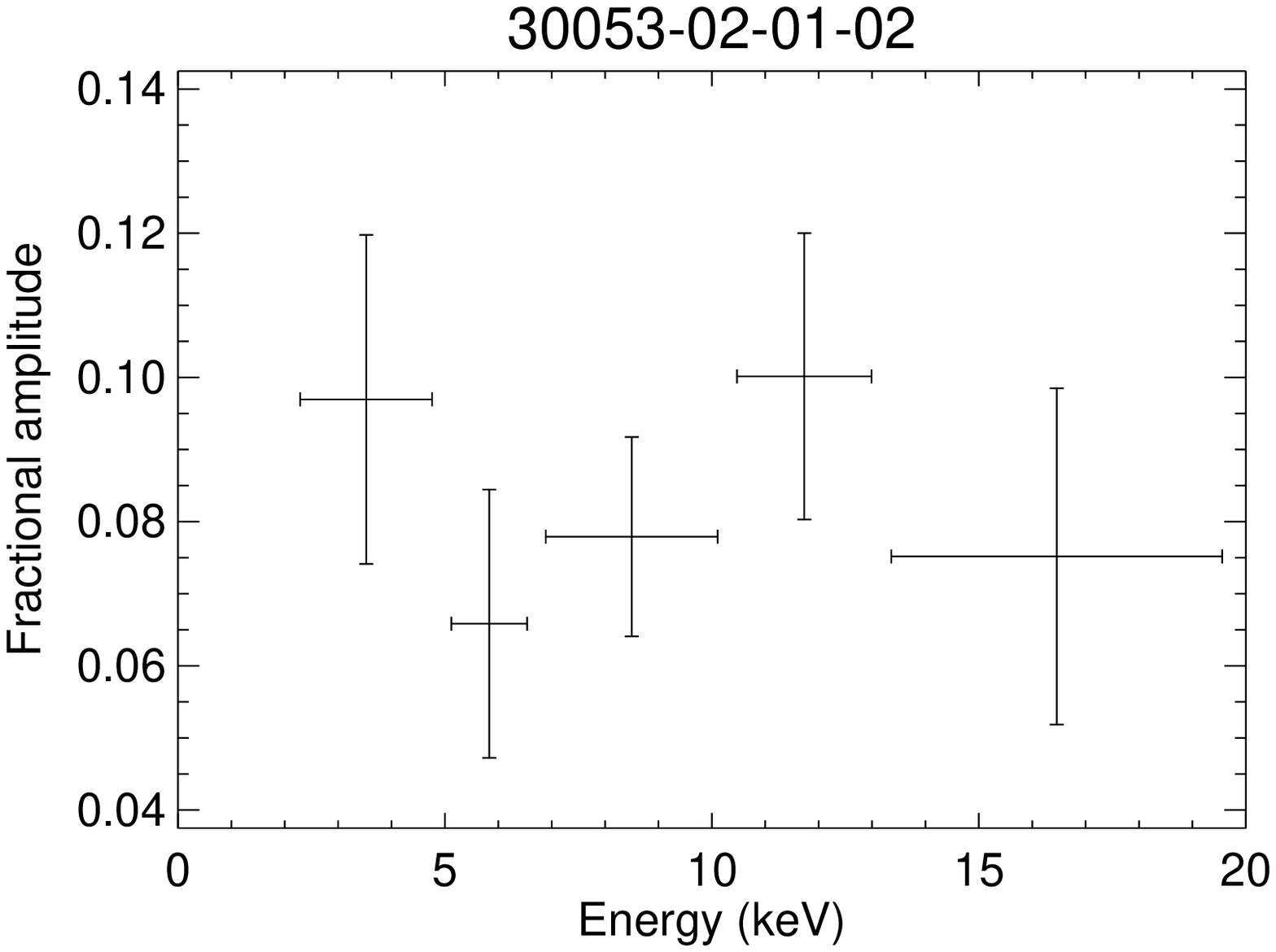} &
\includegraphics[width=0.25\textheight]{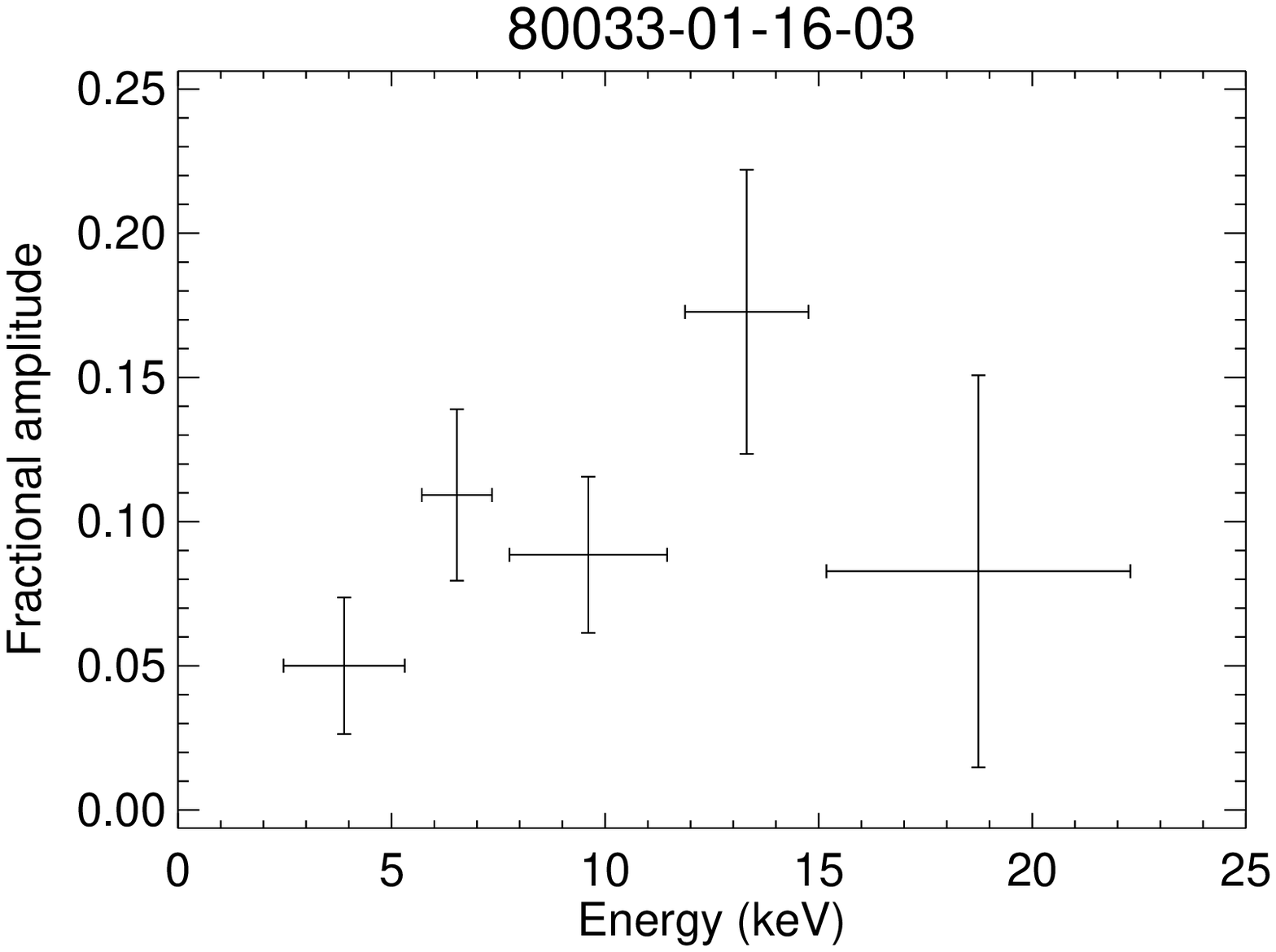} &
\includegraphics[width=0.25\textheight]{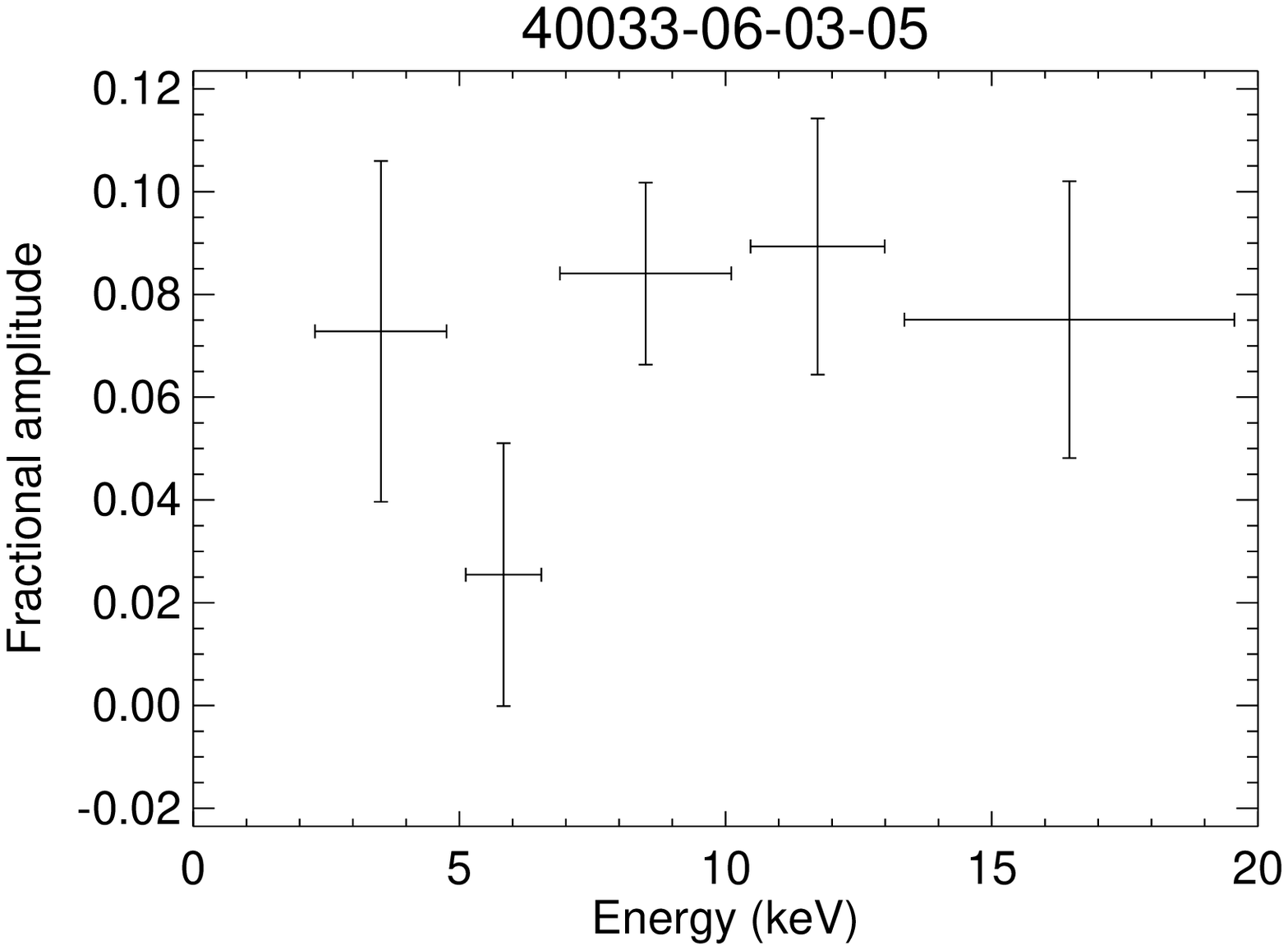} \\
\end{tabular}
\caption{Amplitude variation of burst oscillation with energy. The different columns represent the different sources and the different rows denote the different amplitude evolution with energy. Here we have chosen a representative sample to highlight the amplitude energy behavior across sources. The burst IDs considered for this figure are identified at the top of each panel. \label{fig02}}
\end{figure*}

\begin{figure}
\centering
\includegraphics[width=0.32\textheight]{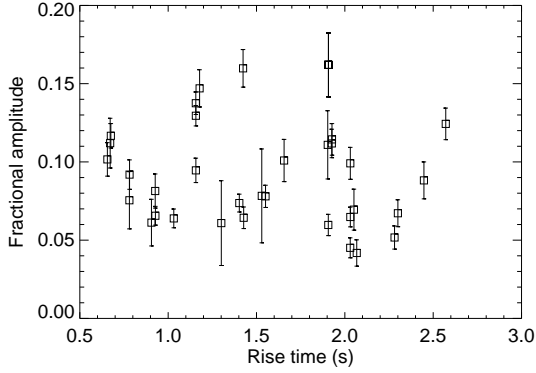}
\caption{Variation of total fractional amplitude with rise time of the burst. The sample of bursts considered for this figure comprises of bursts that exhibit tail oscillations only. The total fractional amplitude was computed by epoch-folding the entire interval during which the oscillation was detected to be significant (see \S~\ref{Observation}).  \label{fig3}}
\end{figure}

\begin{figure*}
\centering
\begin{tabular}{ccc}
\includegraphics[width=0.25\textheight]{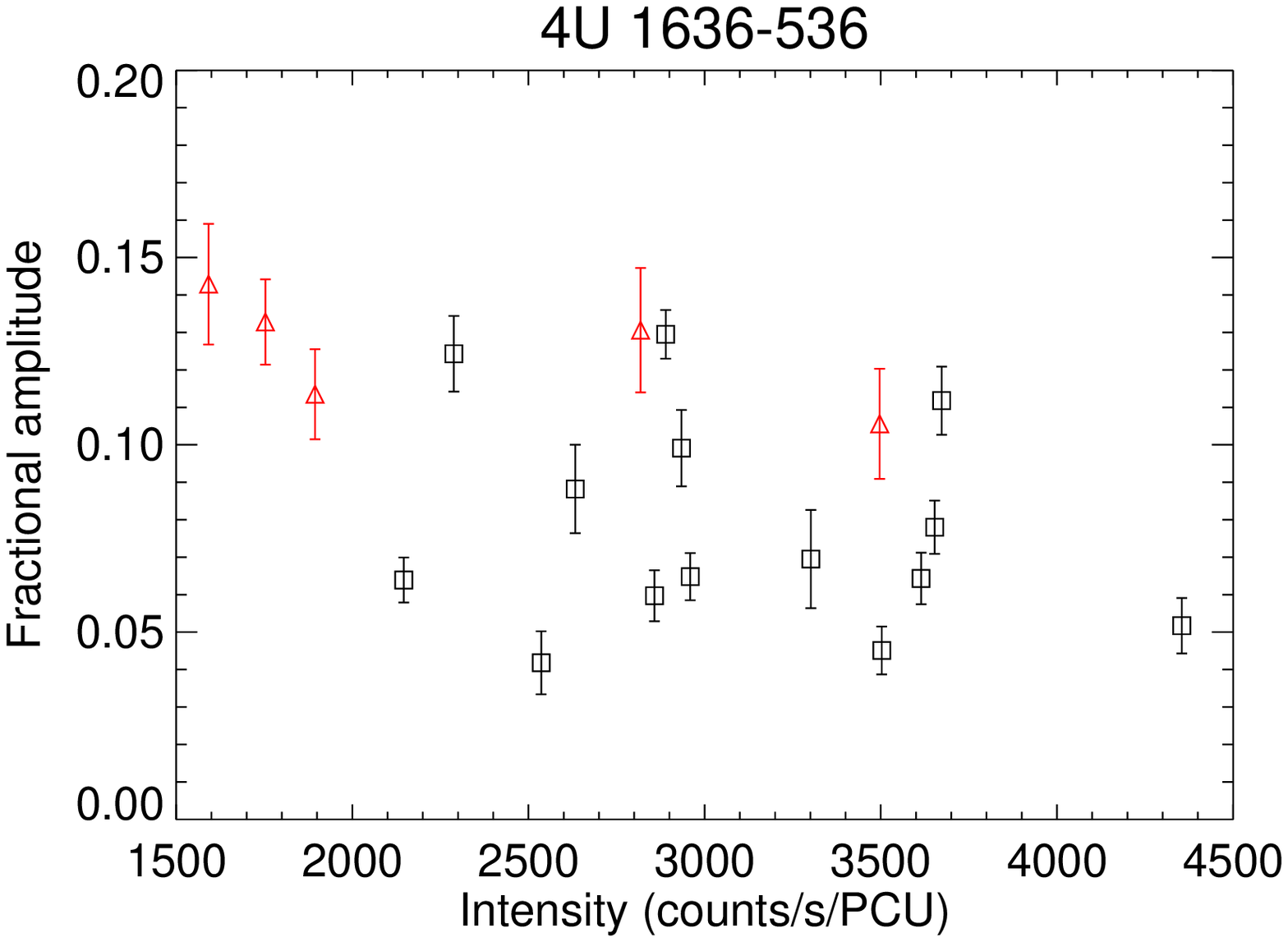} &
\includegraphics[width=0.25\textheight]{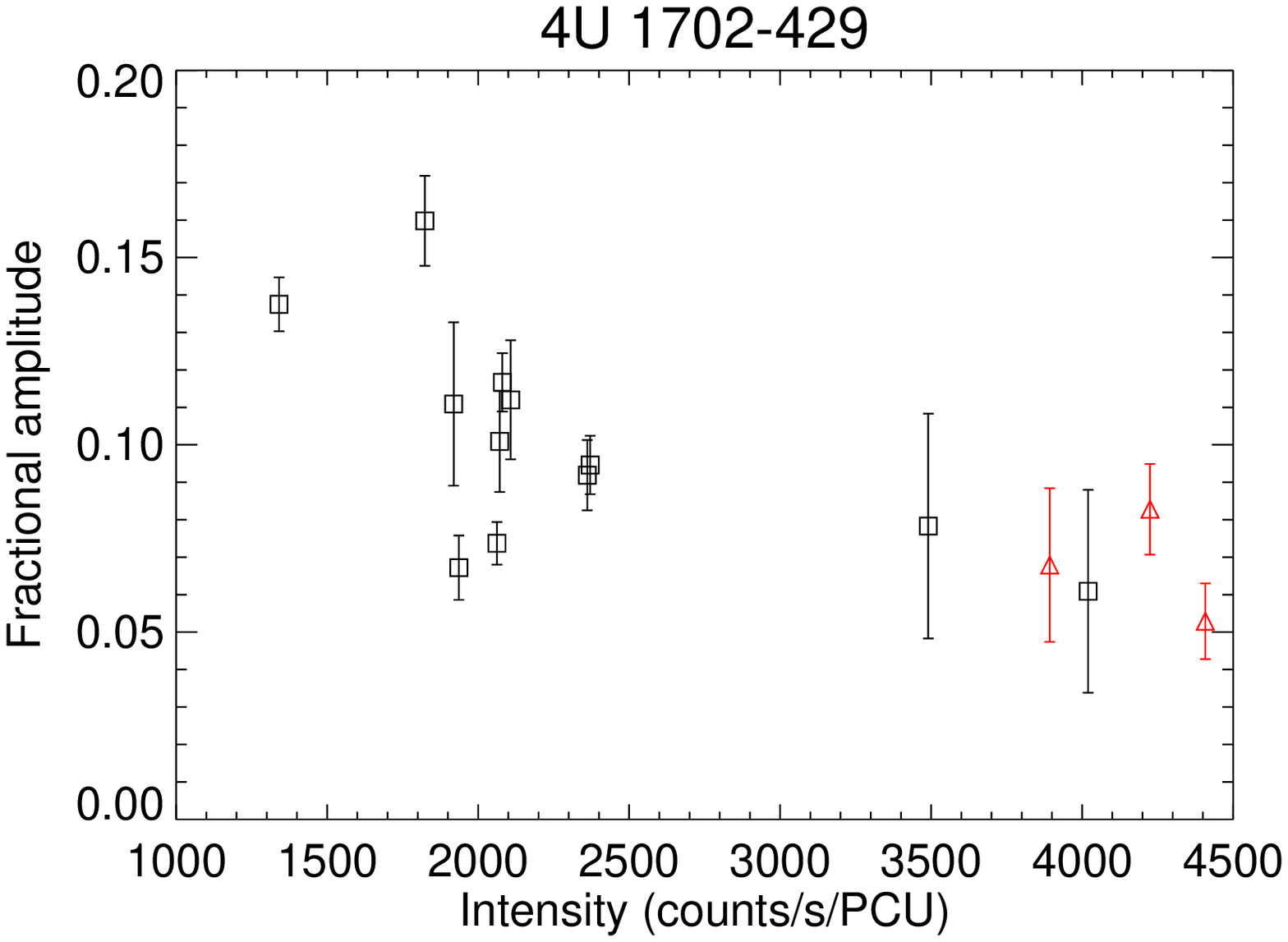} &
\includegraphics[width=0.25\textheight]{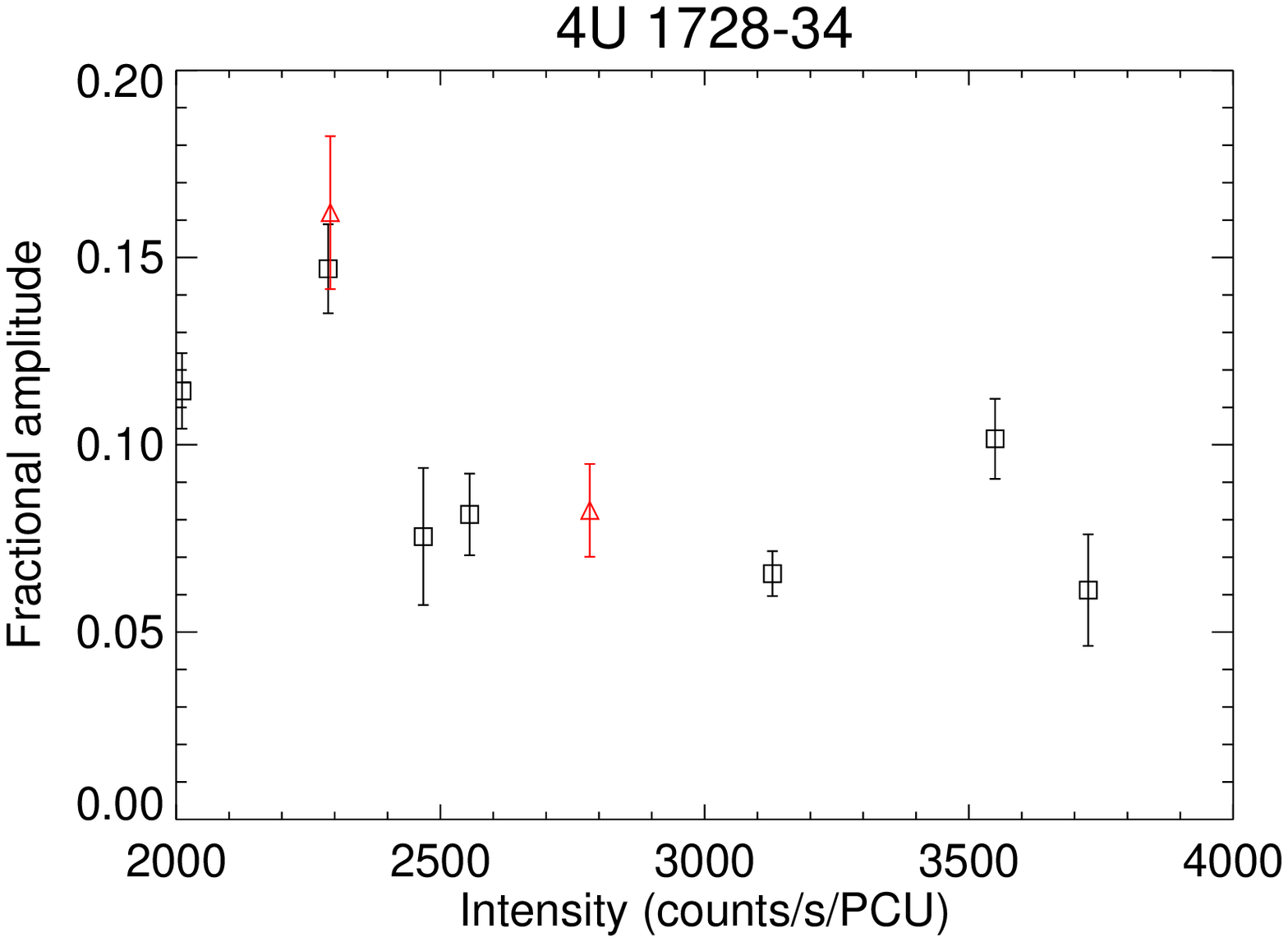} \\
\end{tabular}
\caption{Variation of fractional amplitude with the average intensity in the entire 2-60 keV PCA energy band during the interval where oscillation is present for the three sources 4U 1636--536, 4U 1702--42 and 4U 1728--34. Here the black squares denote the oscillation during the burst decay phase whereas the red triangles denote oscillation during the rising phase. \label{fig4}}
\end{figure*}

\begin{figure*}
\centering
\begin{tabular}{ccc}
\includegraphics[width=0.25\textheight]{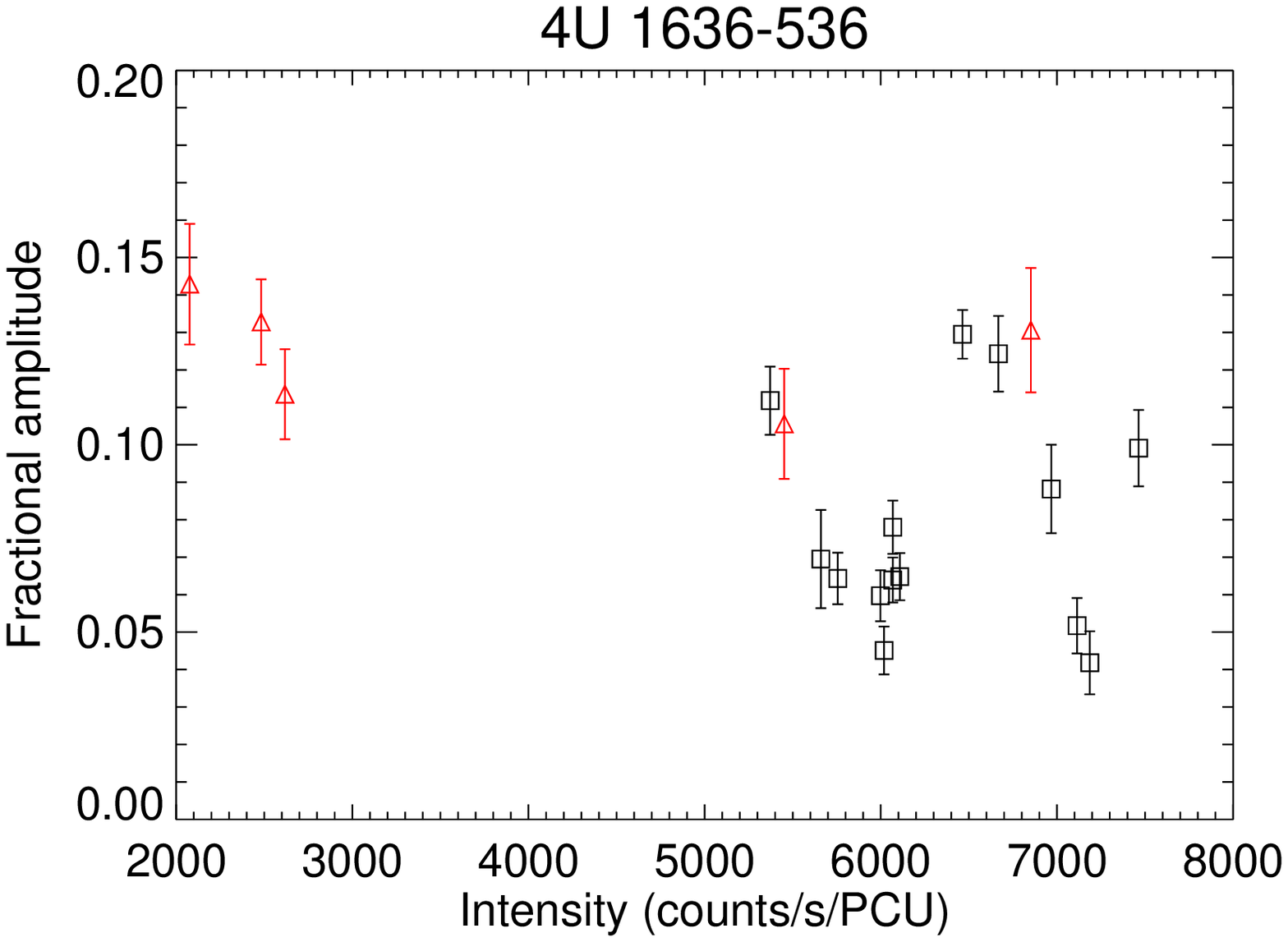} &
\includegraphics[width=0.25\textheight]{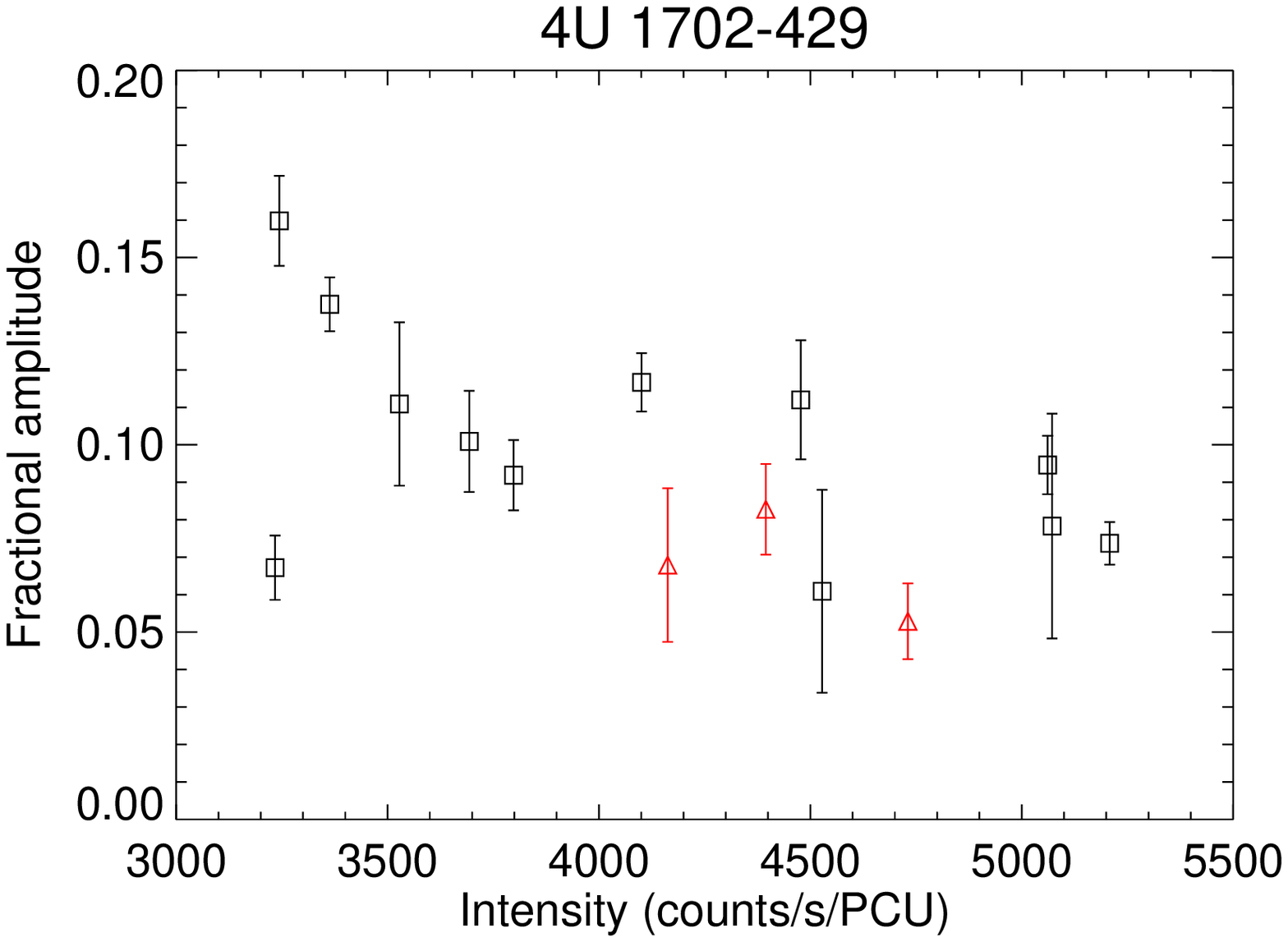} &
\includegraphics[width=0.25\textheight]{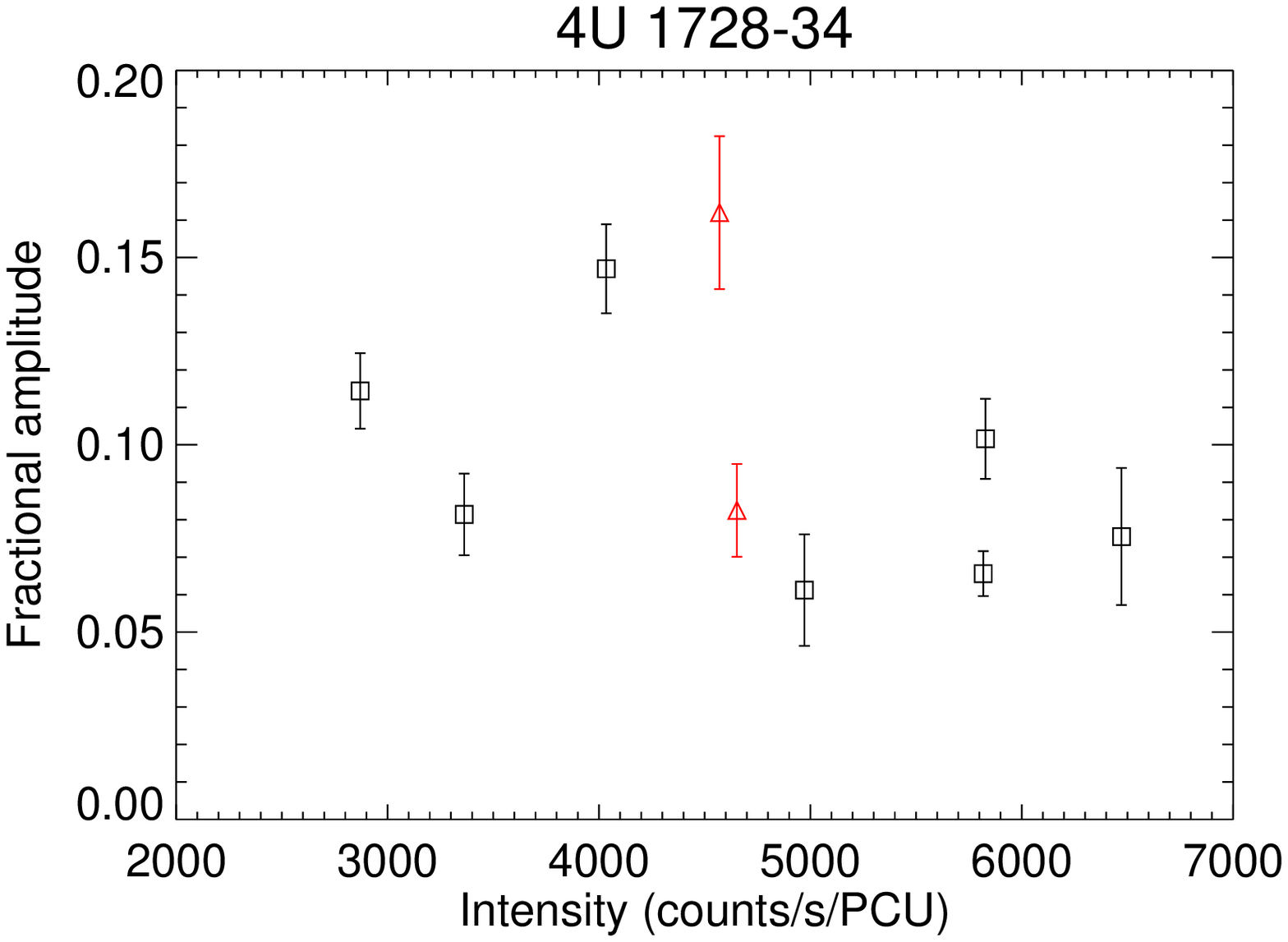} \\
\end{tabular}
\caption{Variation of fractional amplitude with the peak burst intensity in the entire 2-60 keV PCA energy band for the three sources 4U 1636--536, 4U 1702--42 and 4U 1728--34. Here the black squares denote the oscillation during the burst decay phase whereas the red triangles denote oscillation during the rising phase. \label{fig5}}
\end{figure*}

\begin{figure}
\centering
\includegraphics[width=0.35\textheight]{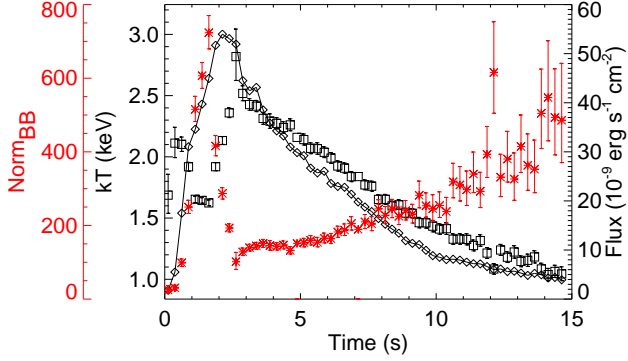}
\caption{Variation of spectral parameters during a typical burst. The solid line with diamond symbols shows the flux each 0.25 s segment. The right hand y-axis corresponds to the burst flux. The black square symbols and the red asterisk symbols displays the variation in the blackbody temperature and normalization respectively. The first red left hand y-axis corresponds to the blackbody normalization and the subsequent left hand y-axis corresponds to the blackbody temperature. \label{fig6}}
\end{figure}

\begin{figure}
\centering
\includegraphics[width=0.32\textheight]{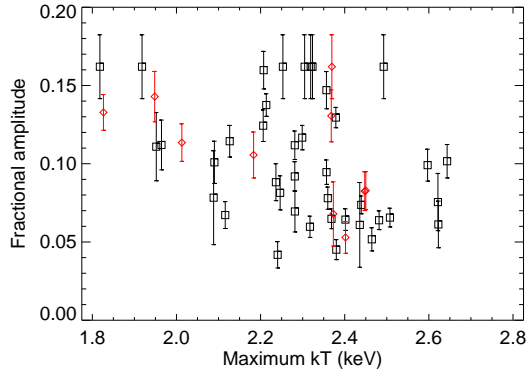}
\caption{Variation of fractional amplitude with the maximum blackbody temperature during the interval where oscillation is present. The red diamonds correspond to the oscillation during the rising phase whereas the black squares represnts the oscillation during the burst decay phsae.\label{fig7}}
\end{figure}

\begin{figure*}
\centering
\begin{tabular}{ccc}
\textbf{4U 1636--536} & \textbf{4U 1702--42} & \textbf{4U 1728--34} \\
\includegraphics[width=0.25\textheight]{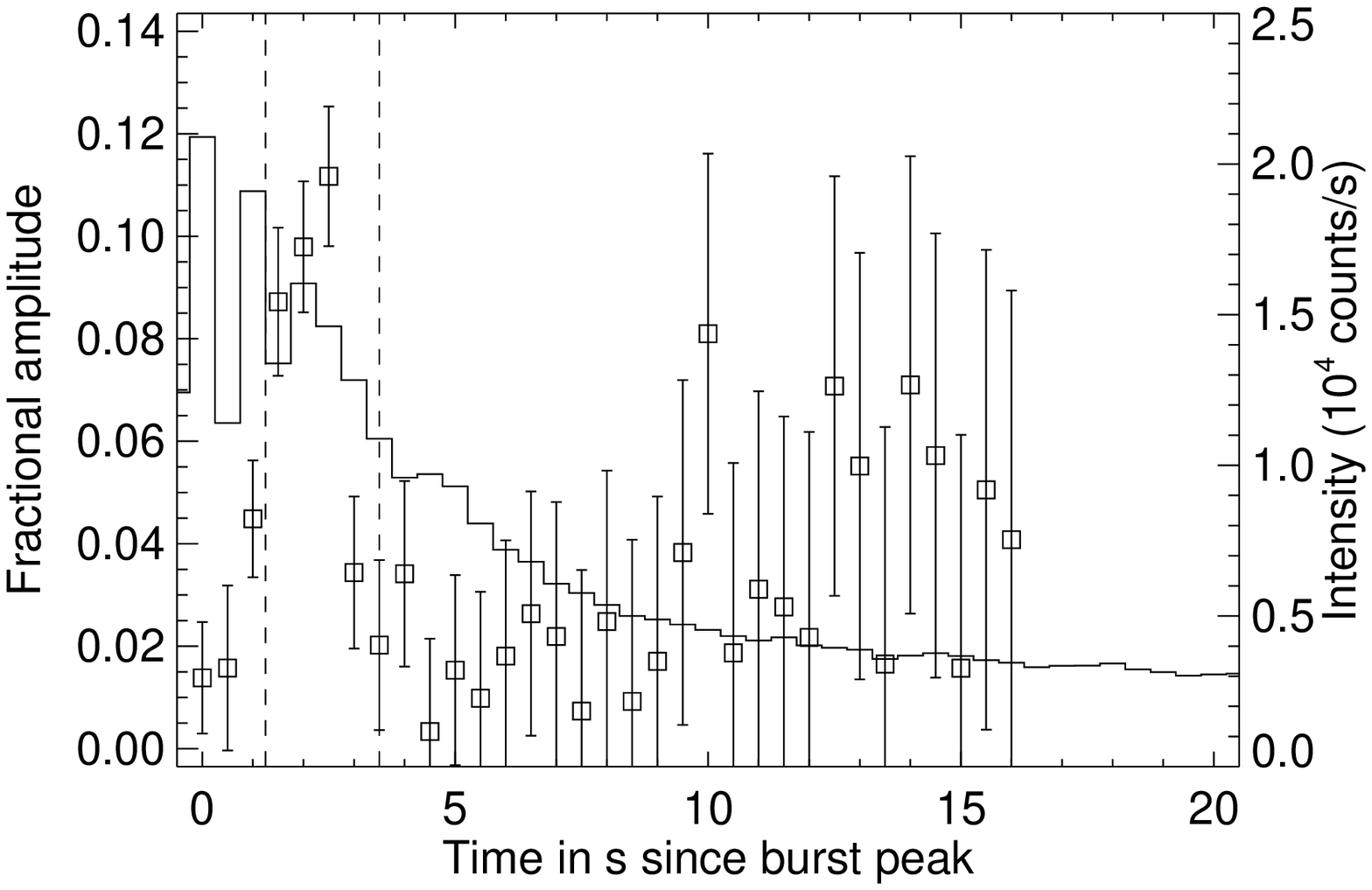} &
\includegraphics[width=0.25\textheight]{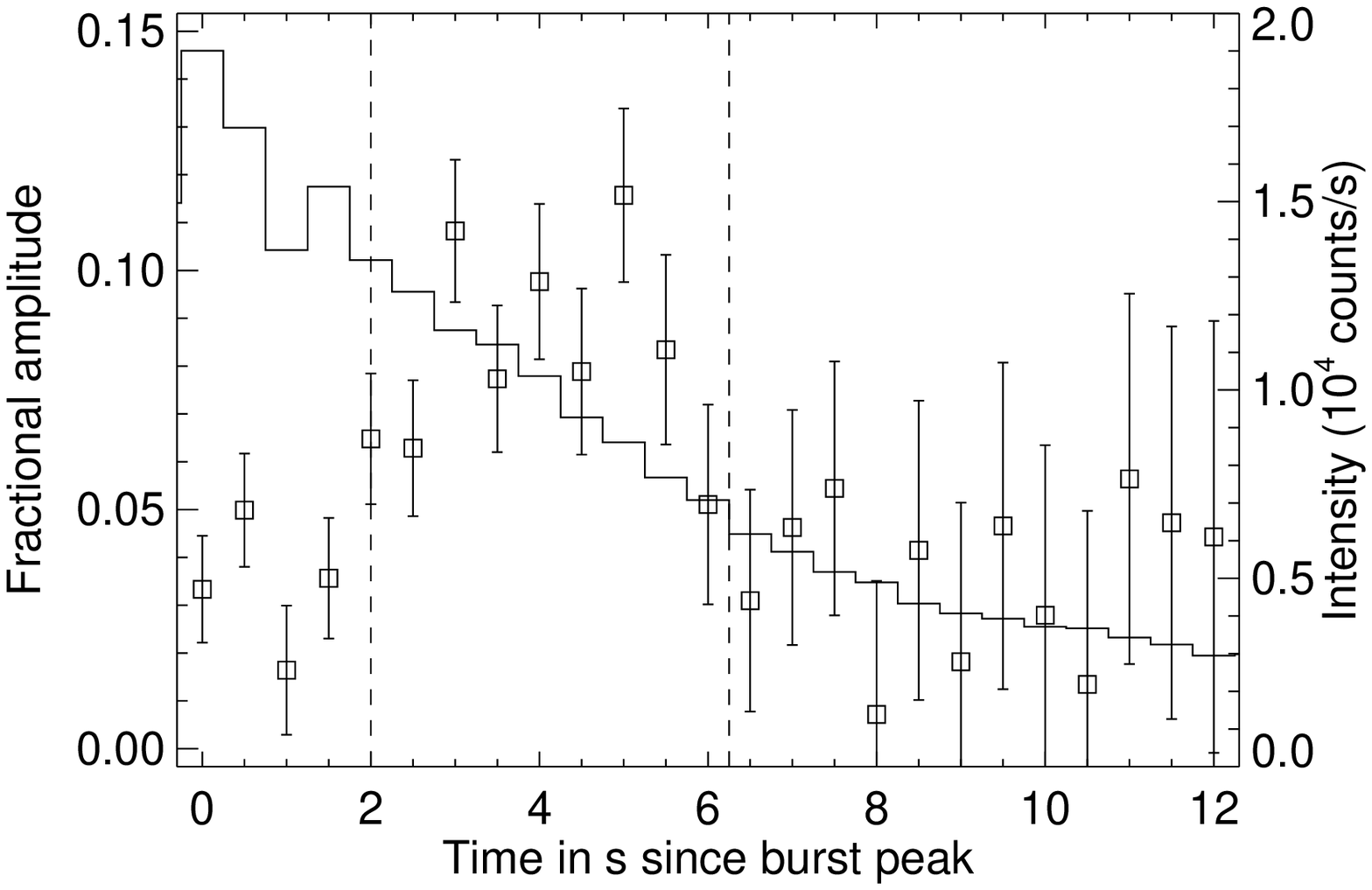} &
\includegraphics[width=0.25\textheight]{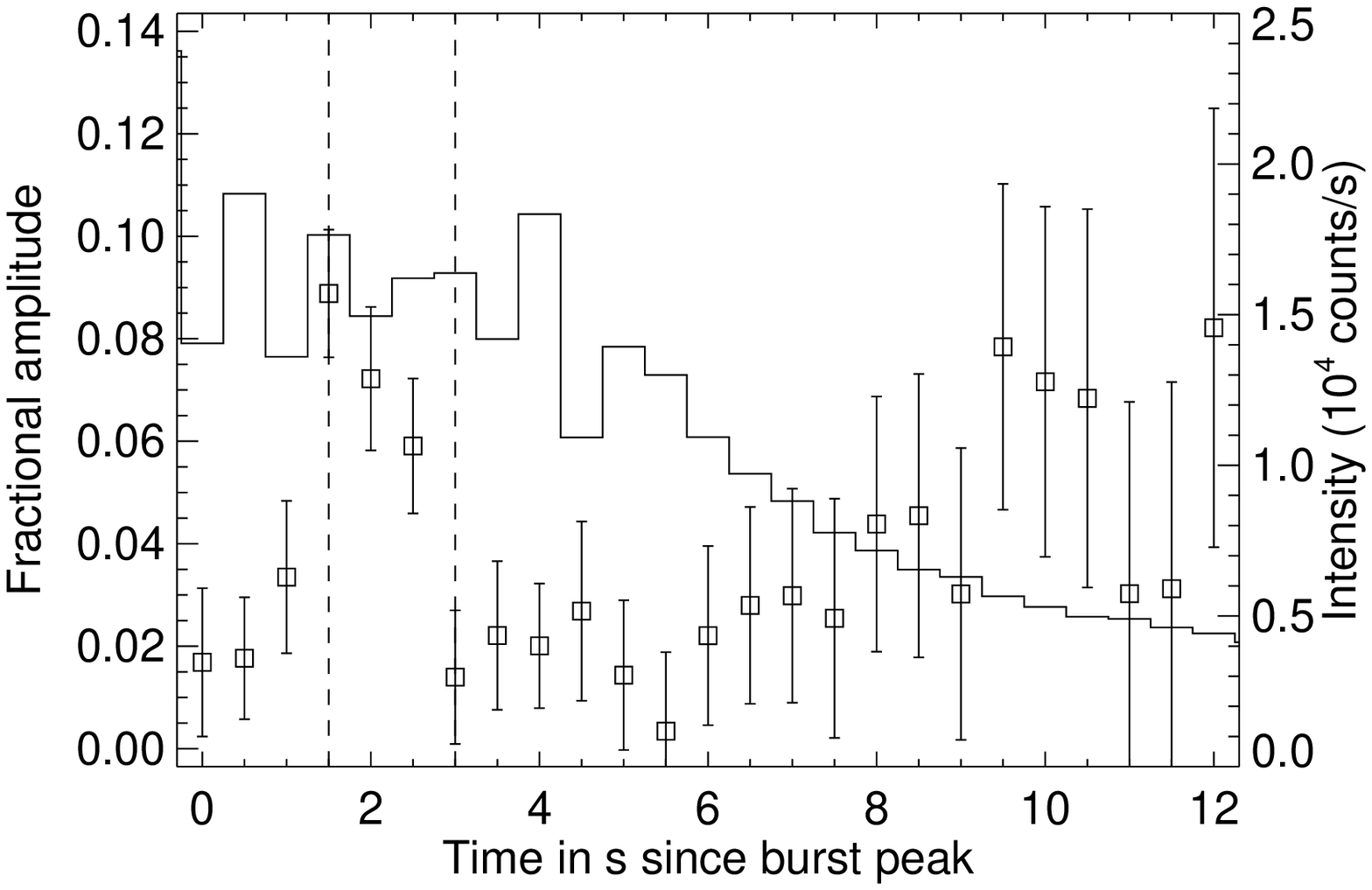} \\
\end{tabular}
\caption{Temporal variation of burst oscillation amplitude for three sources. The dashed lines denote the time interval during which the oscillation was detected to be significant. \label{fig08}}
\end{figure*}

In the rotating hot-spot model, the emission from the hot-spot is modulated by Doppler effect due to a rapidly spinning neutron star. This is supposed to manifest as a lag of the soft energy photons of the burst oscillations. However, we observed mixed behavior in variations of the phase lag with energy: The first row in Fig.~\ref{fig01} shows the representative cases where there is no lag observed for three different sources, the second row corresponds to the case where the phase lag varies non-monotonically with energy for the same sources. The third and the fourth rows correspond to those for which hard and soft lags, respectively, are observed for bursts from these sources. For each source, the slope of the phase lag variation with energy (fitted with a linear model) exhibits both positive and negative values and is uncorrelated with the spin or other stellar properties. For the oscillations in the decaying phase of 4U 1636--536, we detected four cases with negative slope of the phase lag, that is, tentative soft lag among the 14 selected burst decay oscillations; for 4U 1702--429 bursts, we detected six instances with negative slope among 12 burst decay oscillations. For 4U 1728--34, there were four instances with negative slope among 7 burst decay oscillations. For the other sources as well, the slopes of the phase lags were both positive and negative irrespective of the stellar properties. It is important to note here that a negative or a positive slope does not always necessitate a soft or a hard lag as in certain cases the phase lag evolution is well described by a constant function when compared against a linear model. 

An alternative argument to explain the phase lag variation invoked Comptonization of burst photons via scattering in the hot corona \citep{Ford1999, Muno2003}. Additionally, the interaction of the burst photons with the rotating accretion disk was also suggested to have effects though in this case it was harder to explain the observed trends \citep{Muno2003}. However, \citet{Artigue2013} has modeled the theoretically expected phase evolutions, compared them with observations from four bursts from 4U 1636--536. In their model, the source related parameters (mass, radius, viewing latitude, and distance) were kept constant, while the spot latitude, spot radius, and color temperature were free to constrain. They inferred that after taking into account the range of the parameter space, the data is statistically consistent with the predictions of the rotating hot-spot model. In this scenario, the deviation of the data from the model can be explained by the statistical fluctuations even for a correct model possibly due to folding the profile over seconds and thereby smearing the signal. The observed soft leads are also then consistent with the statistical fluctuations particularly considering the errors on the phase lag data. We have focused on extending the observational results, particularly without any averaging, and demonstrated that there is a significant number of cases where soft lead or no lead is present in the phase lag variation of the burst oscillations. In the light of these results it needs to be checked whether the statistical fluctuations around the model predictions can still explain the variation in the evolution of phase lag with energy. In the rotating hot-spot model, the effects of the gravity on the oscillation profile could be investigated to put constraints on the neutron star equation of state (Strohmayer 2000). Such a procedure for the rotation powered pulsars is one of the major science goals of NICER. Hence, the consistency of the observed phase lag results with hot-spot model needs to be examined before it can reliably be employed to estimate the neutron star parameters.

The temporal evolution of the oscillation amplitude during the decay phase of the burst generally exhibits an increase following the burst peak, reaches the maximum, and then decreases steadily towards the end of the tail. The origin of the burst oscillations during the decay phase of thermonuclear bursts is still a debated issue. One of the popular models, that is the surface wave modes in the neutron star ocean was unable to fully explain the frequency drifts and the corresponding oscillation amplitudes. Another scenario that has been recently proposed by \citet{Mahmoodifar2016} is the cooling wake models. In particular, two types of cooling mechanisms have been proposed; the canonical cooling model and the asymmetric cooling model. The temporal evolution of the oscillation amplitude is distinctly different between these two scenarios as shown in Fig. 6 of \citet{Mahmoodifar2016}. In these models the brightness asymmetries are observed when after the flame spreading the fuel burns off and then cools down. In the symmetric canonical cooling model, each patch on the engulfed neutron star ocean cools down at a similar rate and manner. In the asymmetric cooling model, the patches ignited earlier cools faster. \citet{Mahmoodifar2016} concluded that the canonical cooling model cannot reproduce the observed amplitudes during the burst decay unlike the asymmetric cooling model. If the flame speed is sufficiently low the canonical model can in principle reproduce the observed amplitudes but this requires longer rise time. Hence, a correlation between the rise time and oscillation amplitude of the bursts would then be expected in the canonical cooling scenario. We did not observe any significant correlation between the rise time and amplitude, therefore, our results further strengthens the argument for the asymmetric cooling model which is in concurrence with inferences reached by \citet{Mahmoodifar2016}. Moreover from our results we saw that the temporal evolution of the oscillation amplitude during the decay phase of the bursts exhibits a behavior very similar to that predicted in the asymmetric cooling model, the maximum of the amplitude occurring a few seconds after the burst peak.

To further probe the physical components influencing the burst oscillations, we performed spectral studies and determined energetic properties of events corresponding to the burst oscillations. We detected no significant correlation between the average or the maximum temperature during oscillation with the oscillation amplitude. This implies that the strength of oscillation is strongly affected also by factors other the blackbody temperature, that can generate strong brightness contrast. The peak flux and the fluence of the burst were also found to be uncorrelated with amplitude of the corresponding oscillation. We conclude that the energetics of the burst is undoubtedly important but is not the sole contributor towards generation of the brightness asymmetries. The relation between the oscillation amplitude and the average oscillation intensity for individual three sources shows a marginal anti-correlation, which can be attributed to the fact that stronger emission from the whole surface will diminish the brightness contrast of any asymmetry that is formed in the neutron star ocean. 

The temporal evolution of the burst spectral parameters also revealed some interesting behavior. We observed that, particularly during the burst decay, the blackbody normalization generally follows an increasing trend. It has been previously observed by \citet{Zhang2013, Zhang2016} that during the burst decay phase the blackbody radius shows a constant value when the oscillation is present and following the oscillation phase the blackbody radius shows an increasing trend. On the other hand for the bursts where oscillation has not been detected, the blackbody radius usually shows a constant or a decreasing trend. Our result agrees perfectly with these observations as in almost all cases for our selected bursts (all of which exhibit oscillation), that is, we observe an increasing blackbody radius. We also observed comparatively short rise times for our sample of bursts (Table.~\ref{bpartiming}) which was also reported by \citet{Zhang2016} for bursts with oscillation. This phenomenon was attributed to scenario that when the cooling wake initiates near the pole a burst with oscillation is observed and when cooling wake begins near the equator no oscillation is observed during a burst. When the asymmetry begins at a higher latitude the flame front there and hence the cooling there progresses at a relatively slower rate thereby exhibiting a longer lasting asymmetry. Here it should be noted that the last 10 cases that we have presented in the tables show burst oscillation during their rising phase for comparison purposes. Both the blackbody temperature and the blackbody radius show a steady increase during the rising phase of the bursts which can be explained by the spreading of burning front on the surface of the highly rotating star. The oscillation properties during the rise are also in agreement with the expanding hot-spot model which was also pointed out in previous literature \citep{Strohmayer1997, Bhattacharyya2005, Chakraborty2014}. Finally, the convexity indicating the ignition latitude showed both positive and negative values for our sample of bursts, which hints that for the bursts with oscillation during the tail the thermonuclear burning possibly initiates at the higher latitudes which is consistent with our spectral analysis results.

\begin{table*}
\caption{Parameters of burst oscillations \label{bpartiming}}
\begin{tabular}{cccccccccc}
\toprule
Source & Spin     & Burst  & Burst time & T$_{dur}$ & Burst rise & T$_{osc}$ & I$_{peak}$ & I$_{osc}$   & A \\
       & (Hz)     & ID$^1$ & UTC        & (s)$^2$ & time (s)   & (s)$^3$ & (cnts/s)$^4$   & (cnts/s)$^5$     & (\%)$^6$ \\
\hline
4U 1636--536  &  581  &  30053-02-01-02  &  1998-08-20T03:26:36  &  13.53  &   2.03  &   2.00  &     7464.00  &     2934.20  &   8.59$\pm$ 0.82   \\
4U 1636--536  &  581  &  40028-01-06-00  &  1999-06-10T05:11:28  &  17.41  &   1.03  &  10.25  &     6068.00  &     2146.05  &   5.47$\pm$ 0.48   \\
4U 1636--536  &  581  &  40030-03-04-00  &  1999-06-19T16:57:43  &  17.78  &   1.41  &   2.25  &     5372.00  &     3672.78  &  10.06$\pm$ 0.78   \\
4U 1636--536  &  581  &  40031-01-01-06  &  1999-06-21T18:05:36  &  14.30  &   2.05  &   1.25  &     5660.00  &     3301.60  &   6.19$\pm$ 1.11   \\
4U 1636--536  &  581  &  40028-01-15-00  &  2000-06-15T03:49:32  &  16.66  &   1.16  &   5.00  &     6464.00  &     2889.52  &  11.20$\pm$ 0.53   \\
4U 1636--536  &  581  &  40028-01-18-00  &  2000-08-09T08:41:36  &  14.66  &   2.03  &   6.25  &     6108.00  &     2959.20  &   5.65$\pm$ 0.52   \\
4U 1636--536  &  581  &  50030-02-04-00  &  2001-01-28T02:24:16  &  15.66  &   1.53  &   3.75  &     6068.00  &     3653.00  &   7.00$\pm$ 0.60   \\
4U 1636--536  &  581  &  50030-02-05-01  &  2001-02-01T21:00:33  &  14.78  &   1.91  &   6.00  &     5998.00  &     2857.75  &   5.07$\pm$ 0.54   \\
4U 1636--536  &  581  &  60032-01-02-00G  &  2001-06-15T00:03:28  &  13.42  &   2.55  &   5.75  &     6668.00  &     2288.17  &  11.29$\pm$ 0.87   \\
4U 1636--536  &  581  &  60032-01-06-01  &  2001-08-28T06:15:47  &  14.91  &   1.41  &   3.75  &     5756.00  &     3614.53  &   5.87$\pm$ 0.61   \\
4U 1636--536  &  581  &  60032-01-14-01  &  2001-11-01T06:58:27  &  16.78  &   1.78  &   6.00  &     6018.67  &     3502.83  &   4.12$\pm$ 0.56   \\
4U 1636--536  &  581  &  60032-01-20-01  &  2002-01-09T09:49:41  &  14.91  &   2.28  &   3.25  &     7114.67  &     4354.46  &   4.90$\pm$ 0.68   \\
4U 1636--536  &  581  &  60032-05-06-00  &  2002-01-14T07:05:36  &  11.15  &   2.03  &   4.75  &     7186.67  &     2536.21  &   3.85$\pm$ 0.75   \\
4U 1636--536  &  581  &  91024-01-42-00  &  2005-05-26T07:04:32  &  13.18  &   2.43  &   3.75  &     6968.00  &     2632.80  &   7.94$\pm$ 1.01   \\
4U 1702--429  &  329  &  20084-02-01-00  &  1997-07-19T18:24:32  &  14.40  &   1.41  &   7.75  &     5208.00  &     2062.35  &   6.75$\pm$ 0.50   \\
4U 1702--429  &  329  &  20084-02-01-02  &  1997-07-30T11:43:28  &  14.53  &   1.16  &   9.50  &     3363.20  &     1340.67  &  11.53$\pm$ 0.56   \\
4U 1702--429  &  329  &  80033-01-01-07  &  2004-01-18T21:10:06  &  16.17  &   0.55  &   1.50  &     4477.33  &     2107.56  &  10.50$\pm$ 1.45   \\
4U 1702--429  &  329  &  80033-01-01-04  &  2004-01-19T23:46:24  &  15.03  &   1.28  &   0.25  &     4528.00  &     4020.00  &   5.86$\pm$ 2.57   \\
4U 1702--429  &  329  &  80033-01-04-00  &  2004-02-29T01:07:28  &  15.44  &   1.32  &   6.00  &     3244.00  &     1823.42  &  13.72$\pm$ 0.96   \\
4U 1702--429  &  329  &  80033-01-07-02  &  2004-04-08T21:37:36  &  16.30  &   2.30  &   4.50  &     3234.00  &     1936.06  &   6.14$\pm$ 0.76   \\
4U 1702--429  &  329  &  80033-01-09-00  &  2004-04-09T21:13:20  &  15.41  &   0.78  &   3.00  &     3798.00  &     2361.50  &   8.54$\pm$ 0.84   \\
4U 1702--429  &  329  &  80033-01-14-00  &  2004-04-14T17:41:20  &  14.66  &   1.66  &   2.50  &     3693.33  &     2071.47  &   8.98$\pm$ 1.14   \\
4U 1702--429  &  329  &  80033-01-16-03  &  2004-04-16T01:15:42  &  15.53  &   1.91  &   1.00  &     3528.00  &     1919.00  &   9.97$\pm$ 1.86   \\
4U 1702--429  &  329  &  80033-01-16-00  &  2004-04-16T20:09:20  &  13.15  &   0.66  &   4.25  &     4100.80  &     2080.09  &  10.55$\pm$ 0.67   \\
4U 1702--429  &  329  &  80033-01-16-08  &  2004-04-17T02:27:28  &  12.28  &   1.16  &   6.00  &     5061.33  &     2370.89  &   8.64$\pm$ 0.69   \\
4U 1702--429  &  329  &  91023-02-02-00  &  2005-09-17T10:11:28  &  14.03  &   1.53  &   0.25  &     5072.00  &     3490.67  &   7.48$\pm$ 2.80   \\
4U 1728--34  &  363  &  20083-01-02-000  &  1997-09-22T05:59:48  &  15.78  &  1.91  &   3.75  &     2870.40  &     2011.09  &   9.08$\pm$0.73   \\
4U 1728--34  &  363  &  40033-06-03-020  &  1999-01-31T21:57:20  &  12.93  &   0.68  &   1.50  &     5828.80  &     3549.67  &   9.04$\pm$ 0.91   \\
4U 1728--34  &  363  &  40033-06-03-05  &  1999-02-04T21:54:24  &  11.78  &   0.66  &   0.75  &     6472.00  &     2467.73  &   6.51$\pm$ 1.48   \\
4U 1728--34  &  363  &  40019-03-02-00  &  1999-08-19T15:10:24  &  15.16  &   1.16  &   5.50  &     4034.67  &     2287.64  &  10.28$\pm$ 0.73   \\
4U 1728--34  &  363  &  50030-03-08-02  &  2001-10-18T03:12:32  &  15.66  &   0.91  &   5.75  &     5818.00  &     3128.39  &   6.01$\pm$ 0.53   \\
4U 1728--34  &  363  &  50030-03-08-00  &  2001-10-18T09:14:24  &  13.43  &   0.91  &   0.75  &     4972.00  &     3726.00  &   5.69$\pm$ 1.33   \\
4U 1728--34  &  363  &  90406-01-01-00  &  2004-03-12T01:29:20  &  16.78  &   1.03  &   4.50  &     3362.67  &     2555.26  &   6.31$\pm$ 0.76   \\
KS 1731--260  &  541  &  10416-01-01-00  &  1996-07-14T03:36:16  &  14.03  &  2.91  &   2.00  &     3880.00  &     2786.10  &   4.70$\pm$0.85   \\
KS 1731--260  &  541  &  30061-01-04-02  &  1999-02-27T16:48:32  &  14.30  &  1.78  &   7.50  &     3056.00  &     1641.39  &   5.52$\pm$0.57   \\
MXB 1658--298  &  567  &  40050-04-04-00  &  1999-04-14T11:16:16  &  10.28  &  0.78  &   3.50  &    1536.00  &      459.71  &  9.81$\pm$1.57   \\
4U 1608--52  &  620  &  30062-01-01-00  &  1998-03-27T13:54:24  &  16.61  &  1.74  &   0.75  &     14292.80  &     2798.13  &  5.93$\pm$1.38   \\
4U 1608--52  &  620  &  70059-03-01-000  &  2002-09-12T03:54:24  &  25.93  &  2.30  &   7.25  &    14621.33  &     5057.06  &  6.82$\pm$0.43   \\
SAX J1750.8--2900  &  601  &  60035-01-02-02  &  2001-04-12T13:40:32  &  10.43  &  1.82  &   1.00  &   4090.00  &   1545.25  &  9.42$\pm$1.80   \\
4U 1915--053  &  270  &  30066-01-03-03  &  1998-08-01T18:01:47  &  17.18  &  1.53  &   1.50  &    1252.80  &    854.13  &  10.76$\pm$1.76   \\
\hline
4U 1636--536  &  581  &  10088-01-07-02  &  1996-12-28T23:26:24  &  7.28  &  2.41  &  2.50  &  2076.80  &  1591.92  &  10.15$\pm$1.01   \\
4U 1636--536  &  581  &  30053-02-02-00  &  1998-08-20T05:11:07  &  8.43  &  1.68  &  2.75  &  2617.60  &  1894.18  &   9.12$\pm$0.88   \\
4U 1636--536  &  581  &  40028-01-02-00  &  1999-02-27T07:29:20  &  11.03  &  1.91  &  1.00  &  6852.00  &  2818.25  &  11.24$\pm$1.34   \\
4U 1636--536  &  581  &  40028-01-08-00  &  1999-06-18T21:25:36  &  11.91  &  1.78  &  1.00  &  5452.00  &  3496.75  &   9.16$\pm$1.20   \\
4U 1636--536  &  581  &  60032-05-03-00  &  2002-01-12T21:36:32  &  17.17  &  3.17  &  4.00  &  2482.67  &  1753.17  &  11.94$\pm$0.98   \\
4U 1702--429  &  329  &  80033-01-04-02  &  2004-02-29T06:07:28  &  15.03  &  1.03  &  0.50  &  4162.67  &  3892.67  &  6.35$\pm$1.86   \\
4U 1702--429  &  329  &  80033-01-06-00  &  2004-03-02T06:42:24  &  13.78  &  1.78  &  1.75  &  4730.67  &  4408.19  &  5.00$\pm$0.93   \\
4U 1702--429  &  329  &  80033-01-13-000  &  2004-04-13T18:01:20  &  14.28  &  1.66  &  1.25  &  4394.67  &  4224.87  &  7.87$\pm$1.12   \\
4U 1728--34   &  363  &  20083-01-01-01  &  1997-09-19T12:29:44  &  12.40  &  1.90  &  1.50  &  4652.80  &  2782.67  &  7.01$\pm$0.98   \\
4U 1728--34   &  363  &  50030-03-09-01  &  2001-10-27T23:33:56  &  15.91  &  1.91  &  1.00  &  4570.67  &  2291.67  &  14.35$\pm$1.71   \\

\hline
\hline
\end{tabular} 
\\
\begin{flushleft}
$^1$ Since each of the selected RXTE observation sample contain only one thermonuclear burst, the RXTE observation identification number is used as burst identifier. \\
$^2$ The duration of the burst, obtained by considering the time from rise to the instant when the intensity in 10\% the peak intensity.
$^3$ The interval during which the oscillation was found to be significant according to the chosen criterion (\S~\ref{Observation}). \\
$^4$ Peak intensity/PCU of the burst over the entire 2-60 keV energy band computed from light curve with time resolution of 0.125 s. \\
$^5$ Average intensity/PCU during the time when oscillation is present (T$_{osc}$) again computed over the entire 2-60 keV energy band from light curve with time resolution of 0.125 s. \\
$^6$ The time averaged fractional amplitude of oscillation calculated over the interval of significant oscillation (\S~\ref{Observation}). \\
The data following the horizontal line represents the cases for which burst  oscillation has been detected during the rising phase.
\end{flushleft}
\end{table*}

\begin{table*}
\caption{Fit parameters of the phase lags with energy \label{bparslope}} 
\centering
\begin{tabular}{ccccccc}
\toprule
Spin & Burst   & \multicolumn{2}{c}{Constant} &  \multicolumn{3}{c}{Linear} \\
(Hz) & ID      & Level ($10^{-2}$) & $\chi^2 (4)^a$ & Ordinate & slope ($10^{-3}$) & $\chi^2 (3)^a$\\
\hline
581 &  30053-02-01-02 & -0.03 &   3.24 &  -0.0704 &   7.83 &   0.21 \\
581 &  40028-01-06-00 &  0.64 &   0.84 &  -0.0101 &   1.35 &   0.66 \\
581 &  40030-03-04-00 & -0.93 &   4.22 &   0.0593 &  -6.14 &   0.62 \\
581 &  40031-01-01-06 &  0.10 &   4.23 &  -0.0344 &   3.92 &   3.78 \\
581 &  40028-01-15-00 &  0.51 &   4.16 &  -0.0290 &   2.92 &   1.17 \\
581 &  40028-01-18-00 &  0.74 &   6.61 &  -0.0363 &   3.84 &   5.09 \\
581 &  50030-02-04-00 &  0.93 &   5.84 &  -0.0680 &   6.87 &   0.43 \\
581 &  50030-02-05-01 &  1.17 &  10.33 &  -0.0666 &   6.68 &   6.49 \\
581 & 60032-01-02-00G & -0.75 &   4.66 &   0.0502 &  -5.42 &   0.82 \\
581 &  60032-01-06-01 &  2.56 &  16.34 &  -0.0764 &   8.55 &  10.38 \\
581 &  60032-01-14-01 & -0.68 &   5.38 &  -0.0145 &   0.69 &   5.36 \\
581 &  60032-01-20-01 &  0.30 &   1.98 &   0.0074 &  -0.38 &   1.98 \\
581 &  60032-05-06-00 & -2.44 &  10.62 &   0.1275 & -14.54 &   2.21 \\
581 &  91024-01-42-00 &  0.36 &   0.76 &  -0.0178 &   1.99 &   0.67 \\
329 &  20084-02-01-00 &  1.06 &  10.05 &  -0.0417 &   4.64 &   7.55 \\
329 &  20084-02-01-02 & -0.30 &   0.92 &   0.0119 &  -1.46 &   0.45 \\
329 &  80033-01-01-07 & -2.56 &   8.70 &   0.1320 & -12.87 &   1.55 \\
329 &  80033-01-01-04 &  0.68 &   2.25 &   0.0040 &   0.24 &   2.25 \\
329 &  80033-01-04-00 &  0.67 &   2.07 &  -0.0235 &   2.74 &   1.00 \\
329 &  80033-01-07-02 & -0.76 &   8.12 &   0.0655 &  -5.85 &   5.70 \\
329 &  80033-01-09-00 & -0.57 &   0.94 &   0.0011 &  -0.71 &   0.90 \\
329 &  80033-01-14-00 &  1.12 &   7.39 &  -0.0203 &   2.83 &   7.00 \\
329 &  80033-01-16-03 & -0.98 &   0.59 &   0.0158 &  -2.72 &   0.46 \\
329 &  80033-01-16-00 & -0.11 &   6.94 &   0.0490 &  -4.42 &   4.37 \\
329 &  80033-01-16-08 &  0.51 &   3.88 &  -0.0425 &   4.77 &   0.79 \\
329 &  91023-02-02-00 &  0.51 &   1.84 &  -0.0369 &   3.49 &   1.41 \\
363 & 20083-01-02-000 & -0.15 &   2.53 &   0.0589 &  -6.54 &   0.70 \\
363 & 40033-06-03-020 &  0.25 &   3.31 &   0.0539 &  -6.23 &   2.42 \\
363 &  40033-06-03-05 &  0.73 &   1.61 &  -0.0669 &   7.29 &   0.48 \\
363 &  40019-03-02-00 &  0.73 &   4.29 &  -0.0437 &   4.47 &   1.57 \\
363 &  50030-03-08-02 &  0.18 &   6.45 &  -0.0269 &   2.32 &   5.79 \\
363 &  50030-03-08-00 & -0.24 &   1.02 &   0.0102 &  -1.37 &   1.00 \\
363 &  90406-01-01-00 &  0.41 &   2.59 &   0.0303 &  -2.13 &   2.44 \\
541 &  10416-01-01-00 &  1.79 &   3.99 &  -0.0353 &   4.81 &   3.21 \\
541 &  30061-01-04-02 &  0.45 &   3.14 &  -0.0276 &   3.53 &   1.96 \\
567 &  40050-04-04-00 &  1.73 &   1.38 &  -0.0441 &   5.51 &   0.38 \\
620 &  30062-01-01-00 & -1.01 &   3.50 &   0.0581 &  -6.32 &   2.77 \\
620 & 70059-03-01-000 &  0.72 &   2.77 &  -0.0152 &   1.90 &   2.22 \\
601 &  60035-01-02-02 &  0.84 &   1.65 &  -0.0049 &   1.03 &   1.62 \\
270 &  30066-01-03-03 & -0.61 &   5.76 &   0.1225 & -12.51 &   2.78 \\
\hline
581 &  10088-01-07-02 & -0.40 &   2.30 &  -0.0054 &   0.14 &   2.30 \\
581 &  30053-02-02-00 &  1.20 &  10.72 &  -0.0758 &   8.06 &   6.92 \\
581 &  40028-01-02-00 &  0.39 &   0.56 &  -0.0086 &   1.30 &   0.50 \\
581 &  40028-01-08-00 &  0.77 &   1.94 &  -0.0160 &   2.02 &   1.74 \\
581 &  60032-05-03-00 &  0.45 &   0.66 &  -0.0057 &   1.02 &   0.56 \\
329 &  80033-01-04-02 &  0.59 &   9.88 &  -0.1716 &  17.71 &   2.40 \\
329 &  80033-01-06-00 &  0.80 &   4.04 &  -0.0762 &  10.36 &   1.16 \\
329 & 80033-01-13-000 &  0.75 &   2.70 &  -0.0434 &   6.13 &   1.57 \\
363 &  20083-01-01-01 &  0.33 &   2.94 &  -0.0586 &   5.95 &   1.77 \\
363 &  50030-03-09-01 &  1.13 &   0.96 &  -0.0292 &   3.64 &   0.37 \\
\hline
\hline
\end{tabular}
\begin{flushleft}
$^a$ The chi-square of the constant and linear fits with the corresponding degrees of freedom presented in the parentheses. \\
As before, the data points following the horizontal line represent those cases where the oscillation has been detected during the rising phase of the burst.
\end{flushleft}
\end{table*}

\clearpage
\begin{table*}
\caption{Spectral parameters of burst oscillations \label{bparspec}}
\centering
\begin{tabular}{ccccc}
\toprule
Burst ID & kT$_{osc}^1$ & Peak flux$^2$ ($10^{-9}$) & Fluence$^3$ ($10^{-7}$) & Convexity \\
       & (keV)      & erg/s/cm$^2$          & erg/cm$^2$ & \\
\hline
30053-02-01-02 &  2.34  $\pm$  0.24 &    62.68 $_{    0.59}^{    0.59}$ &   3.55 &   7.26 \\
40028-01-06-00 &  2.04  $\pm$  0.28 &    51.47 $_{    0.47}^{    0.47}$ &   3.66 &   3.41 \\
40030-03-04-00 &  2.17  $\pm$  0.11 &    47.16 $_{    0.47}^{    0.46}$ &   3.22 &  14.83 \\
40031-01-01-06 &  2.16  $\pm$  0.17 &    49.26 $_{    0.47}^{    0.47}$ &   2.88 &   9.46 \\
40028-01-15-00 &  2.20  $\pm$  0.16 &    51.98 $_{    0.49}^{    0.48}$ &   3.70 &   6.31 \\
40028-01-18-00 &  1.92  $\pm$  0.33 &    51.35 $_{    0.51}^{    0.51}$ &   2.99 &  -0.93 \\
50030-02-04-00 &  2.21  $\pm$  0.13 &    45.70 $_{    0.37}^{    0.36}$ &   3.28 &   5.60 \\
50030-02-05-01 &  1.95  $\pm$  0.32 &    52.19 $_{    0.48}^{    0.47}$ &   3.09 &  -4.78 \\
60032-01-02-00G &  1.97  $\pm$  0.23 &    60.10 $_{   0.60}^{    0.60}$ &   3.81 &  30.36 \\
60032-01-06-01 &  2.18  $\pm$  0.23 &    46.56 $_{    0.44}^{    0.43}$ &   3.12 &   1.29 \\
60032-01-14-01 &  2.06  $\pm$  0.27 &    49.66 $_{    0.51}^{    0.50}$ &   3.66 &  17.65 \\
60032-01-20-01 &  2.34  $\pm$  0.11 &    59.04 $_{    0.58}^{    0.58}$ &   4.06 &  21.70 \\
60032-05-06-00 &  1.86  $\pm$  0.27 &    60.67 $_{    0.58}^{    0.57}$ &   3.23 &  27.04 \\
91024-01-42-00 &  2.01  $\pm$  0.18 &    56.84 $_{    0.57}^{    0.57}$ &   3.50 &  29.24 \\
20084-02-01-00 &  1.90  $\pm$  0.40 &    51.64 $_{    0.49}^{    0.48}$ &   3.08 &  28.20 \\
20084-02-01-02 &  1.53  $\pm$  0.42 &    28.92 $_{    0.29}^{    0.28}$ &   1.70 &  10.53 \\
80033-01-01-07 &  1.85  $\pm$  0.16 &    41.29 $_{    0.43}^{    0.44}$ &   2.61 & -18.26 \\
80033-01-01-04 &  2.44  $\pm$  0.03 &    42.70 $_{    0.45}^{    0.44}$ &   2.65 &  11.44 \\
80033-01-04-00 &  1.78  $\pm$  0.31 &    26.16 $_{    0.42}^{    0.42}$ &   1.59 &  10.05 \\
80033-01-07-02 &  1.69  $\pm$  0.30 &    29.72 $_{    0.32}^{    0.32}$ &   2.18 &  25.73 \\
80033-01-09-00 &  2.00  $\pm$  0.28 &    34.52 $_{    0.35}^{    0.34}$ &   2.08 &   6.65 \\
80033-01-14-00 &  1.89  $\pm$  0.20 &    34.21 $_{    0.41}^{    0.40}$ &   1.96 &  29.13 \\
80033-01-16-03 &  1.84  $\pm$  0.16 &    31.01 $_{    0.38}^{    0.37}$ &   1.82 &  24.52 \\
80033-01-16-00 &  1.96  $\pm$  0.28 &    32.82 $_{    0.32}^{    0.31}$ &   2.00 &   7.24 \\
80033-01-16-08 &  1.84  $\pm$  0.37 &    46.46 $_{    0.46}^{    0.46}$ &   2.44 &  13.98 \\
91023-02-02-00 &  2.09  $\pm$  0.02 &    46.70 $_{    0.47}^{    0.47}$ &   2.90 &  32.07 \\
20083-01-02-000 &  1.88 $\pm$   0.21 &    23.52 $_{   0.26}^{    0.26}$ &   1.52 &  19.17 \\
40033-06-03-020 &  2.60 $\pm$   0.07 &    55.01 $_{   0.55}^{    0.54}$ &   3.31 &  -1.85 \\
40033-06-03-05 &  2.60  $\pm$  0.04 &    59.64 $_{    0.60}^{    0.59}$ &   3.46 &  10.96 \\
40019-03-02-00 &  2.08  $\pm$  0.26 &    30.58 $_{    0.41}^{    0.42}$ &   1.74 &   8.76 \\
50030-03-08-02 &  2.22  $\pm$  0.30 &    53.66 $_{    0.54}^{    0.53}$ &   3.48 &   9.34 \\
50030-03-08-00 &  2.59  $\pm$  0.05 &    47.49 $_{    0.45}^{    0.44}$ &   2.74 &   0.98 \\
90406-01-01-00 &  2.12  $\pm$  0.17 &    25.46 $_{    0.36}^{    0.36}$ &   1.87 &   8.64 \\
10416-01-01-00 &  2.19  $\pm$  0.15 &    29.81 $_{    0.30}^{    0.30}$ &   1.95 &  23.74 \\
30061-01-04-02 &  2.05  $\pm$  0.23 &    34.36 $_{    0.33}^{    0.33}$ &   2.49 &  21.94 \\
40050-04-04-00 &  2.19  $\pm$  0.24 &    10.48 $_{    0.17}^{    0.17}$ &   0.46 &  -6.37 \\
30062-01-01-00 &  1.76  $\pm$  0.08 &   128.10 $_{    1.17}^{    1.16}$ &   9.92 &   6.57 \\
70059-03-01-000 &  2.11 $\pm$   0.19 &   129.69 $_{   1.17}^{    1.17}$ &  14.33 &  -4.53 \\
60035-01-02-02 &  1.82  $\pm$  0.13 &    39.89 $_{    0.39}^{    0.39}$ &   1.93 &  10.00 \\
30066-01-03-03 &  2.13  $\pm$  0.17 &    10.93 $_{    0.18}^{    0.18}$ &   0.71 &  23.55 \\
\hline
10088-01-07-02 &  1.82  $\pm$  0.19 &    13.40 $_{    0.19}^{    0.20}$ &   0.40 &  -5.44 \\
30053-02-02-00 &  1.85  $\pm$  0.18 &    19.99 $_{    0.23}^{    0.24}$ &   0.65 &  13.57 \\
40028-01-02-00 &  2.37  $\pm$  0.03 &    54.44 $_{    0.53}^{    0.54}$ &   2.78 &   9.59 \\
40028-01-08-00 &  1.98  $\pm$  0.28 &    45.38 $_{    0.46}^{    0.44}$ &   2.30 &  14.74 \\
60032-05-03-00 &  1.74  $\pm$  0.10 &    19.58 $_{    0.27}^{    0.28}$ &   1.47 &  19.56 \\
80033-01-04-02 &  2.37  $\pm$  0.03 &    37.79 $_{    0.42}^{    0.42}$ &   2.54 &   5.13 \\
80033-01-06-00 &  2.28  $\pm$  0.14 &    45.06 $_{    0.85}^{    0.85}$ &  10.87 &  21.45 \\
80033-01-13-000 &  2.40  $\pm$  0.09 &    41.26 $_{   0.44}^{    0.44}$ &   2.58 &  15.61 \\
20083-01-01-01 &  2.38  $\pm$  0.09 &    39.75 $_{    0.39}^{    0.39}$ &   2.45 &   8.55 \\
50030-03-09-01 &  2.33  $\pm$  0.06 &    42.19 $_{    0.46}^{    0.46}$ &   2.79 &  11.71 \\
\hline
\hline
\end{tabular}
\\
\begin{flushleft}
$^1$ The blackbody temperature from the time-resolved spectroscopy analysis averaged over the duration during which the burst oscillation is present. The errors presented are usually standard deviation; 1 $\sigma$ statistical errors are presented in certain cases where it is comparatively larger. \\
$^2$ The 3-15 keV flux and its 1 $\sigma$ error corresponding to the burst peak.  \\
$^3$ The 3-15 keV burst fluence integrated over the burst duration. \\ 
The data following the horizontal line represents the cases for which burst  oscillation has been detected during the rising phase only. \\
\end{flushleft}
\end{table*}

\begin{table*}
\caption{Maximum of burst oscillation amplitudes \label{bparosct}}
\centering
\begin{tabular}{ccc}
\toprule
Burst ID & Position relative$^1$ &  Maximum fractional$^2$ \\
         & to burst peak (s)     &  amplitude in \%     \\
\hline
30053-02-01-02  &        1.00  &        7.61  $\pm$        0.97 \\
40028-01-06-00  &        4.50  &       13.20  $\pm$        1.62 \\
40030-03-04-00  &        2.50  &       11.20  $\pm$        1.37 \\
40031-01-01-06  &        3.00  &        6.87  $\pm$        1.43 \\
40028-01-15-00  &        3.50  &       14.23  $\pm$        1.67 \\
40028-01-18-00  &        4.50  &       10.26  $\pm$        1.66 \\
50030-02-04-00  &        1.50  &        7.84  $\pm$        1.53 \\
50030-02-05-01  &        8.50  &        9.78  $\pm$        2.90 \\
60032-01-02-00G  &        7.50  &       15.35  $\pm$        3.38 \\
60032-01-06-01  &        3.00  &        5.41  $\pm$        1.27 \\
60032-01-14-01  &        5.50  &        9.03  $\pm$        1.97 \\
60032-01-20-01  &        3.50  &        7.21  $\pm$        1.48 \\
60032-05-06-00  &        6.00  &       15.74  $\pm$        2.70 \\
91024-01-42-00  &        5.00  &       11.02  $\pm$        2.54 \\
20084-02-01-00  &        7.50  &       10.02  $\pm$        2.19 \\
20084-02-01-02  &        6.00  &       15.89  $\pm$        2.44 \\
80033-01-01-07  &        4.50  &       12.63  $\pm$        1.99 \\
80033-01-01-04  &        0.00  &        6.58  $\pm$        1.31 \\
80033-01-04-00  &        7.00  &       18.18  $\pm$        4.51 \\
80033-01-07-02  &        2.50  &        8.58  $\pm$        1.68 \\
80033-01-09-00  &        3.00  &        8.64  $\pm$        1.69 \\
80033-01-14-00  &        2.00  &       10.55  $\pm$        2.03 \\
80033-01-16-03  &        2.50  &       10.48  $\pm$        2.24 \\
80033-01-16-00  &        5.00  &       11.55  $\pm$        1.81 \\
80033-01-16-08  &        7.50  &       12.35  $\pm$        3.38 \\
91023-02-02-00  &        1.50  &        6.17  $\pm$        1.35 \\
20083-01-02-000  &        1.50  &       11.74  $\pm$        1.70 \\
40033-06-03-020  &        1.50  &        8.88  $\pm$        1.24 \\
40033-06-03-05  &        2.00  &        5.97  $\pm$        0.94 \\
40019-03-02-00  &        3.50  &       17.00  $\pm$        2.99 \\
50030-03-08-02  &        1.00  &        8.13  $\pm$        1.52 \\
50030-03-08-00  &        4.50  &        6.74  $\pm$        1.79 \\
90406-01-01-00  &        3.00  &       14.36  $\pm$        2.41 \\
10416-01-01-00  &        0.00  &        6.20  $\pm$        1.59 \\
30061-01-04-02  &        4.00  &       13.70  $\pm$        1.86 \\
40050-04-04-00  &        2.50  &       15.86  $\pm$        3.88 \\
30062-01-01-00  &        5.50  &       11.00  $\pm$        4.48 \\
70059-03-01-000  &        9.00  &        9.65  $\pm$        1.15 \\
60035-01-02-02  &        3.50  &        9.35  $\pm$        2.14 \\
30066-01-03-03  &        2.50  &        9.31  $\pm$        2.59 \\
\hline
10088-01-07-02  &       -0.50  &       12.84  $\pm$        2.06 \\
30053-02-02-00  &       -1.50  &       37.65  $\pm$        4.52 \\
40028-01-02-00  &       -2.00  &       20.20  $\pm$        3.43 \\
40028-01-08-00  &       -1.50  &       16.85  $\pm$        1.94 \\
60032-05-03-00  &       -3.00  &       40.99  $\pm$       10.25 \\
80033-01-04-02  &       -0.50  &        8.29  $\pm$        1.69 \\
80033-01-06-00  &       -0.50  &        5.40  $\pm$        1.31 \\
80033-01-13-000  &       -0.50  &        5.43  $\pm$        1.37 \\
20083-01-01-01  &       -2.00  &       25.19  $\pm$        5.10 \\
50030-03-09-01  &       -2.00  &       24.48  $\pm$        7.14 \\
\hline
\hline
\end{tabular}
\\
\begin{flushleft}
$^1$ The position of the significant maximum of the time evolution of the burst oscillation relative to the burst peak in seconds. A negative value means that oscillation are detected before the burst peak during the rising phase. \\
$^2$ The maximum fractional amplitude in \% and its corresponding error. The fractional amplitude in this case is computed in each 0.5 s time bin during the burst. Here it should be noted that during the rise the amplitude is relatively less in a couple of cases, this is because of the fine time interval with less statistics; and an averaged amplitude computed during the entire rise would reproduce the usual larger amplitude values reported in literature during the rise.  \\
The data following the horizontal line represents the cases for which burst  oscillation has been detected during the rising phase only. \\
\end{flushleft}
\end{table*}

\acknowledgments

We acknowledge support from the Scientific and Technological Research Council of Turkey (T\"UB\.ITAK, grant no: 115R034). We thank the referee for constructive comments and suggestions which have improved the paper.

{}

\end{document}